\documentclass[]{pasj01}

\usepackage{threeparttable}
\usepackage{multirow}

\begin{document}
\SetRunningHead{Pan et al.}{History of NGC 5430}

\title{Constraints on the Minor Merging and Star Formation History of the Wolf-Rayet Galaxy NGC 5430 Through Observations}

\author{Hsi-An \textsc{Pan}} %
\affil{Department of Physics, Faculty of Science, Hokkaido University, Kita 10 Nishi 8 Kita-ku, Sapporo 060-0810, Japan}
\email{hapan@astro1.sci.hokudai.ac.jp}

\author{Nario \textsc{Kuno}}
\affil{Faculty of Pure and Applied Sciences, University of Tsukuba, 1-1-1 Tennoudai, Tsukuba, Ibaraki 350-8577, Japan}
\author{Kazuo \textsc{Sorai}} %
\affil{Department of Physics, Faculty of Science, Hokkaido University, Kita 10 Nishi 8 Kita-ku, Sapporo 060-0810, Japan}
\and
\author{Michiko \textsc{Umei}}
\affil{Department of Physics, Faculty of Science, Hokkaido University, Kita 10 Nishi 8 Kita-ku, Sapporo 060-0810, Japan}


%


\maketitle

\begin{abstract} 
We used multi-wavelength analysis of the newly observed molecular gas ($^{12}$CO and $^{13}$CO (1-–0)) with interferometer CARMA and archival star formation tracers to constrain the interaction, merging, and star formation history of an off-center minor merger, a three-spiral barred galaxy NGC 5430 and its satellite embedded in the bar. Morphology of the molecular gas in the bar of NGC 5430 shows minimal signs of recent interactions in our resolution. The apparent morphological remnant of the past galaxy interaction is an asymmetric spiral arm, containing more molecular gas and exhibiting higher star formation rate (SFR) surface density than the two primary arms.  Rotation curve analysis suggests that  NGC 5430 and its satellite collided several Gyr ago. History of star formation was constrained by using SFRs that trace different timescales (infrared, radio continuum, and H$\alpha$). The collision  occurred  5 -- 10 Myr ago, triggering a transient off-center starburst of Wolf-Rayet stars at the eastern bar end. In the past, the global SFR during the Wolf-Rayet starburst peaked at 35 M$_{\solar}$ yr$^{-1}$. At present, the merger-driven starburst is rapidly decaying and the current global SFR has decreased to the Galactic value. The SFR will continue to decay as suggested by the present amount of dense gas (traced by HCN (1–-0)). Nonetheless, the global SFR is still dominated by the Wolf-Rayet region rather than the circumnuclear region. Compared with other barred galaxies, the circumnuclear region exhibits particularly low dense gas fraction, low star formation activity and high concentration of gas. Physical properties of the molecular gas are inferred by using the large velocity gradient (LVG) calculations. The initial mass ratio of the NGC 5430 and its satellite are suggested to be in the intermediate ratio range of 7:1-20:1. 
\end{abstract}

\section{Introduction}
Hierarchical formation models suggest that galaxy-galaxy interactions play a key role in the formation, evolution, and morphology of galaxies  (e.g. \cite{Whi78,Kaz08}).
 Minor mergers, i.e., mergers of galaxies with large mass ratio, take place frequently in $z$ $<$ 1, and their rate is significantly higher than that of  major mergers (mergers between galaxies of equal mass).
While major mergers tend to form featureless massive elliptical galaxies, minor mergers are more relevant to disk galaxies, because they exert a relatively small effect on the primary disks (e.g. \cite{Mas05,Cas14}).
Minor mergers help to form bulge and increase disk thickness in disk galaxies through deposition of satellite materials \citep{Shi98,Agu01,Eli06,Hop10}. These mergers also help to form galactic bars by triggering disk instability  \citep{Her95,Ski12}, and probably trigger some low or intermediate luminosity active galactic nuclei (AGNs) \citep{Hop08,Kaz08,Vil08,Map15}.

Galaxy-galaxy interactions and mergers trigger star formation. Observations suggest that galaxy-galaxy interactions increases star formation activity through violent compression of molecular gas \citep{Lam03,Kau04,Lin07,Smi07,Woo07,Kav14}. \citet{Kav14} suggested that about half of the star formation events in the local Universe are triggered by the minor merger processes. The notion of merger-induced star formation is also supported by theoretical studies (e.g. \cite{Cox08,Lot10}).

Observations suggest that off-center mergers  are common and can occur repeatedly in a primary disk (e.g. \cite{Man14,Guo15}). The collision sites are shown as discrete star forming regions in the disks of primary galaxies and contribute significant amount of UV and optical light to the host galaxies  \citep{Wuy12,Guo15}.

In this work, we diagnosed the history of mergers and star formation events in a nearby off-center minor-merger system: NGC 5430 and its satellite galaxy HOLM 569B\footnote{In the Holmberg galaxy catalog of galaxy groups, NGC 5430 is HOLM569.}. The results can help to constrain the star formation properties and history of the high-redshift counterparts that often exhibit higher rate of mergers but cannot be resolved in details.

\begin{figure}
 \begin{center}
  \includegraphics[width=0.35\textwidth]{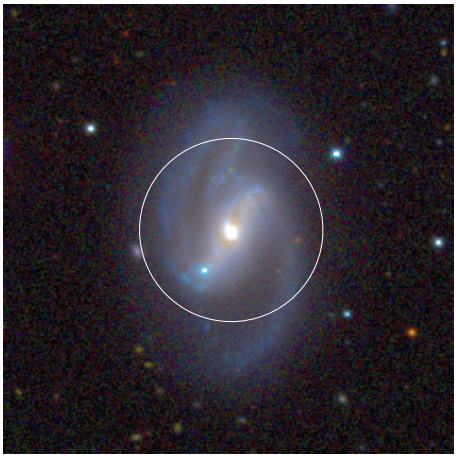} 
 \end{center}
\caption{SDSS \citep{Bai11} $g-r-i$ composite of NGC 5430 system taken from the NED. The white circle indicates the area of CARMA observations. Diameter of the circle is 80$\arcsec$ or $\sim$ 16 kpc.
}
\label{fig_NGC5430SDSSgri}
\end{figure}

A Sloan digital sky survey (SDSS) $g-r-i$ composite image of the primary disk galaxy NGC 5430 is shown in Figure \ref{fig_NGC5430SDSSgri}. NGC 5430 (14$\mathrm{h}$00$\mathrm{m}$45.7$\mathrm{s}$, $+$59$^{\circ}$19$\arcmin$42$\arcsec$) is classified as SB(s)b, located at 42 Mpc, where 1$\arcsec$ $=$ 205 pc. NGC 5430 is a member of a galaxy group \citep{Gel83}. The satellite dwarf galaxy HOLM 569B (14$\mathrm{h}$00$\mathrm{m}$47.3$\mathrm{s}$, $+$59$^{\circ}$19$\arcmin$27$\arcsec$)  is embedded in the eastern bar end, and is shown as a blue spot in Figure \ref{fig_NGC5430SDSSgri}, around 4.5 kpc (22$\arcsec$) from the galaxy center \citep{Keel82}. A 2D Gaussian fit to its $i$-band suggests that the diameter of HOLM 569B is $\sim$ 1 kpc. 

The minor-merger system is manifested by the asymmetric spiral arms and the unusual star formation activity at the eastern bar end. NGC 5340 contains three spiral arms, including two primary arms and one asymmetric arm located at the north of the galactic nucleus, implying a morphology distortion caused by an external effect. The global infrared luminosity of NGC 5430 is not particularly high ($\sim$ 7.5 $\times$ 10$^{10}$ $L_{\solar}$; \cite{San03}) and is inferior to the luminous infrared galaxy (LIRG; $\geq$ 10$^{11}$ $L_{\solar}$). However, the giant HII region  at the eastern bar end is unusually bright, with star formation rate equal to that of the entire Milky Way. Moreover, the giant HII region deviates from the global HII regions luminosity function of the galaxy, suggesting that the bright core is a separate object  \citep{Keel82}. The most massive stars are forming in the giant HII region, including  10$^{4}$-10$^{5}$ OB stars and  Wolf-Rayet stars ($>$ 20 M$_{\solar}$) \citep{Keel82,Fer04}. In the rest of this work, we refer to the eastern bar end (the position of the HOLM 569B and the surrounding 1-kpc-diameter area) as the Wolf-Rayet (W-R) region. The contribution of UV light and the size of the W-R region are similar to those of the giant star forming clumps that are formed through minor mergers in the local and high-redshift Universe \citep{Elm07,Liv12,Guo15}.

The spatial distribution of molecular gas in the NGC 5430 was unknown prior to this work, while single-point observations were performed in a few studies. \citet{Kru90}  observed low density gas in $^{12}$CO (1--0), $^{12}$CO (2--1) and $^{13}$CO (2--1) in the galactic nucleus by using the IRAM 30-m  telescope. Later, \citet{Con97} used the same telescope to measure dense gas tracers HCN (1--0) and CS (3--2), as well as to perform $^{12}$CO (2--1) observations towards the nucleus and the W-R region. No CS (3–-2) emission was detected in both regions, while HCN (1-–0) was detected in the nucleus. The HCN--far infrared (FIR) luminosity relation for this nucleus is the same as for other nearby galaxies; however, the $^{12}$CO (2--1), as well as the $^{12}$CO (1--0) intensities are unusually high relative to the FIR fluxes. The unusual nuclear CO intensities were reported by \citet{Con97} for two other Wolf-Rayet  galaxies. $^{12}$CO (2--1) was detected in the W-R region. The intensity of $^{12}$CO (2--1) in the W-R region was $~$13 times weaker than in the nucleus.

This paper is structured as follows. The mapping observations of $^{12}$CO (1--0) and $^{13}$CO (1--0) are described in Section \ref{sec_obs}. The results of these observations, including the images, morphological and kinematic analyses, are shown in Section \ref{sec_results}. In Section \ref{sec_phy_prop_sf} , we present the physical properties of molecular gas and the history of star formation. In Section \ref{sec_mass_ratio}, we discuss the possible initial mass ratio of NGC 5430 and its satellite galaxy. In Section 6, we summarize our work.

\section{Observation and Data Reduction}
\label{sec_obs}
Molecular gas of NGC 5430 was mapped by using the combined array for research in millimeter-wave astronomy (CARMA). CARMA consists of six 10.4-m-diameter antennas and nine 6.1-m-diameter antennas. Observations were performed on July 30th, August 1st and 3rd of 2013, by using E configuration, with baselines in the 8 -- 66 m range. Observations were performed towards the central 80$\arcsec$ (diameter of $\sim$ 16 kpc) of the galaxy (white circle in Figure \ref{fig_NGC5430SDSSgri}). Two narrow bands (250 MHz) were centered at 115.27 GHz ($^{12}$CO (1-–0)) and 110.20 GHz ($^{13}$CO (1-–0)). The total velocity width of a narrow band was $\sim$ 740 km $^{-1}$, with spectral resolution of $\sim$ 9 km s$^{-1}$. Wide bands (500 MHz) were placed between the narrow bands for the calibration of continuum. Flux, passband, and phase calibrators were mwc349, 1635+381, and 1642+689, respectively. The total observation time was $\sim$ 15 hours, including the target and the calibrators.

Calibration, imaging and deconvolution were performed by using the standard procedures of the MIRIAD package \citep{Sau95}. The final $^{12}$CO (1--0) and $^{13}$CO  (1--0)  clean maps have the spatial resolutions of 7.6$\arcsec$ (1.5 kpc) $\times$ 5.3$\arcsec$ (1.1 kpc) with P.A. $=$ -71.8$^{\circ}$, and 8.1$\arcsec$ (1.7 kpc) $\times$ 5.5$\arcsec$ (1.1 kpc) with P.A. $=$ -68.0$^{\circ}$, respectively. With the velocity resolution binned to 12 km s$^{-1}$, the final spectral cube of $^{12}$CO (1–-0) has the sensitivity of $\sigma_{\mathrm{RMS}}$ $\approx$  11 mJy beam$^{-1}$. Because $^{13}$CO  (1--0) is significantly weaker than $^{12}$CO (1--0), it has been binned to the lower spectral resolution of 25 km s$^{-1}$ for improving the signal-to-noise ratio (S/N). The final $\sigma_{\mathrm{RMS}}$ of $^{13}$CO  (1--0) is 6 mJy beam$^{-1}$. The largest structure that can be detected in these observations is $\sim$ 82$\arcsec$ or $\sim$ 17 kpc. Therefore  missing flux is negligible for this observation.

\section{Results}
\label{sec_results}
\subsection{Galactic Morphology}
\label{sec_morph}
\subsubsection{Channel Maps of CO}
\begin{figure*}
\begin{center}
   \includegraphics[width=0.9\textwidth] {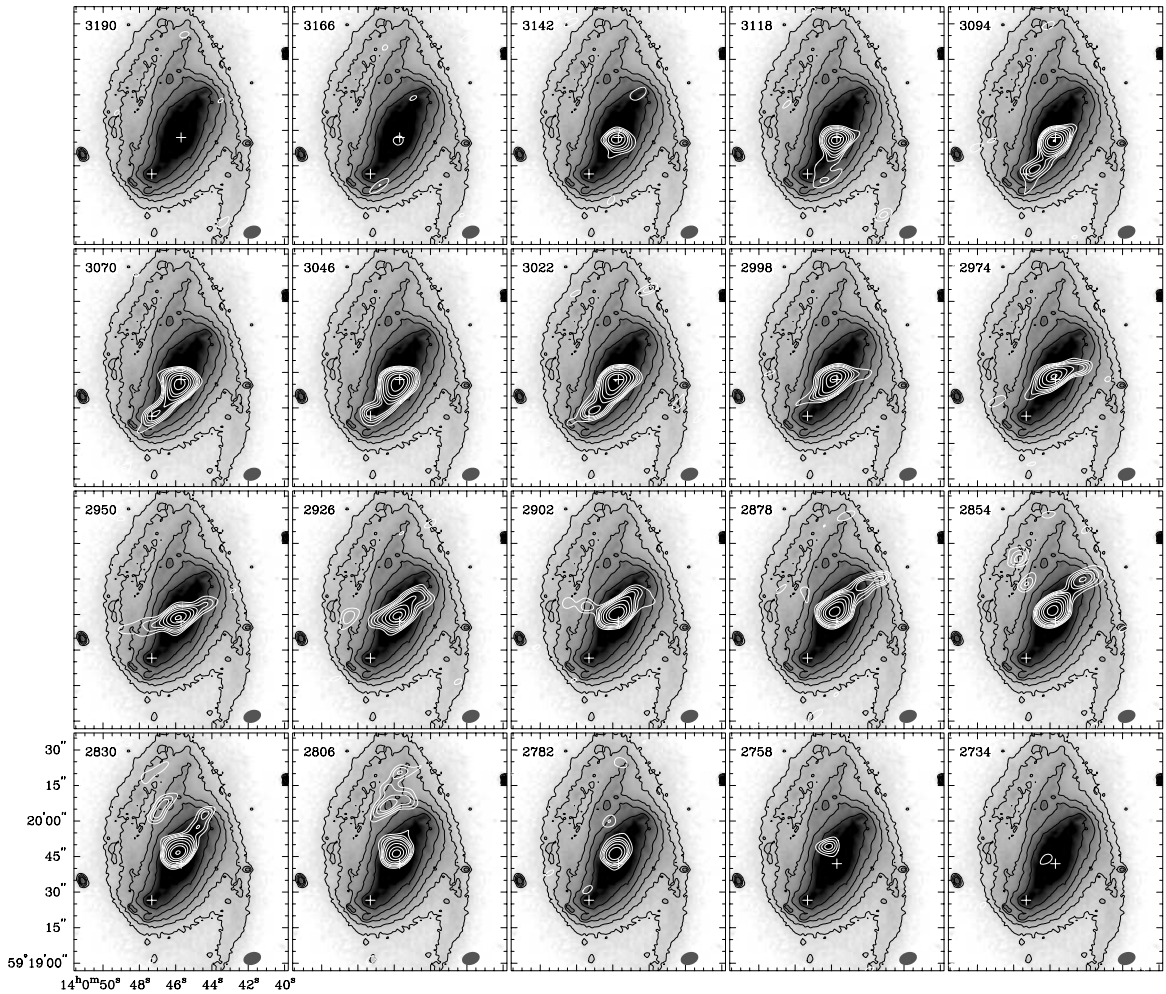}
   \end{center}
\caption{Channel map of $^{12}$CO (1--0) in white contours overlaid on the SDSS $i$-band image \citep{Bai11} in black contours and grey scale. Velocity resolution of the data cube is 12 km s$^{-1}$. The figure presents every other channel to save the space.  The contours of $^{12}$CO are 2, 3, 4, 5, 7, 10, 15, 20, and 30 $\sigma_{\mathrm{RMS}}$, where 1 $\sigma_{\mathrm{RMS}}$ is 11 mJy. The  crosses denote the galactic center (upper one) and the W-R region (lower one), respectively. Beam size of $^{12}$CO (1--0) is 7.6$\arcsec$ $\times$ 5.3$\arcsec$  (1.5 $\times $1.1 kpc) with P.A. $=$ -71.8$^{\circ}$, showing in the lower-right corner of each channel. Velocity of each channel is displayed at the upper-left.
}
 \label{fig_co_ch}
\end{figure*}

The channel map of $^{12}$CO (1--0) (hereafter $^{12}$CO) is delineated in Figure \ref{fig_co_ch} with white contours. The SDSS $i$-band image is also shown, in both greyscale and black contours. The galactic center and the W-R region are marked with crosses, respectively. The $^{12}$CO exhibits a wide range of velocities, from 2730 to 3170 km s$^{-1}$. The strongest emission was generated at the galactic center ($\sim$ 40 $\sigma_{\mathrm{RMS}}$ or $\sim$ 0.44 Jy beam$^{-1}$). Two elongated structures emerge from the center, and were detected with the significance of $\sim$ 10 $\sigma_{\mathrm{RMS}}$  or $\sim$  0.11 Jy beam$^{-1}$. The elongated structures correspond to the bar regions of the optical image. The W-R region was detected with the significance of $\sim$ 10 $\sigma_{\mathrm{RMS}}$ or $\sim$ 0.11 Jy beam$^{-1}$. 

The channel map of $^{12}$CO reveals features that are typical of a barred galaxy. The $^{12}$CO emission from the bar shifts towards the downstream of galactic rotation, known as the offset ridges of the bar. The offset ridges trace the shocks that are associated with the crowded gas orbits, as suggested by theoretical studies \citep{Ath92}. The two offset ridges appear to be spatially and kinematically symmetric with respect to the galactic center. After passing through the offset ridge shocks, the gas moves inwards and accumulates in the central region, yielding high gas concentration at the center  (e.g., \cite{Kun07}). This phenomenon is well-known as bar-driven gas inflow  \citep{Sak99}. The widths of velocity ranges of the galactic center, eastern, and western offset ridges were $\sim$ 440, 180, and 160 km s$^{-1}$, respectively. $^{12}$CO was also detected in the asymmetric arm, emerging in the ~2780 to 2860 km s$^{-1}$ range.  $^{12}$CO was observed in a spot of the north primary arm with velocity in the 2854 -- 2830 km s$^{-1}$ range and with significance of $\sim$ 4 $\sigma_{\mathrm{RMS}}$ or 0.044 Jy beam$^{-1}$.

\begin{figure*}
 \begin{center}
  \includegraphics[width=0.6\textwidth, angle=-90]{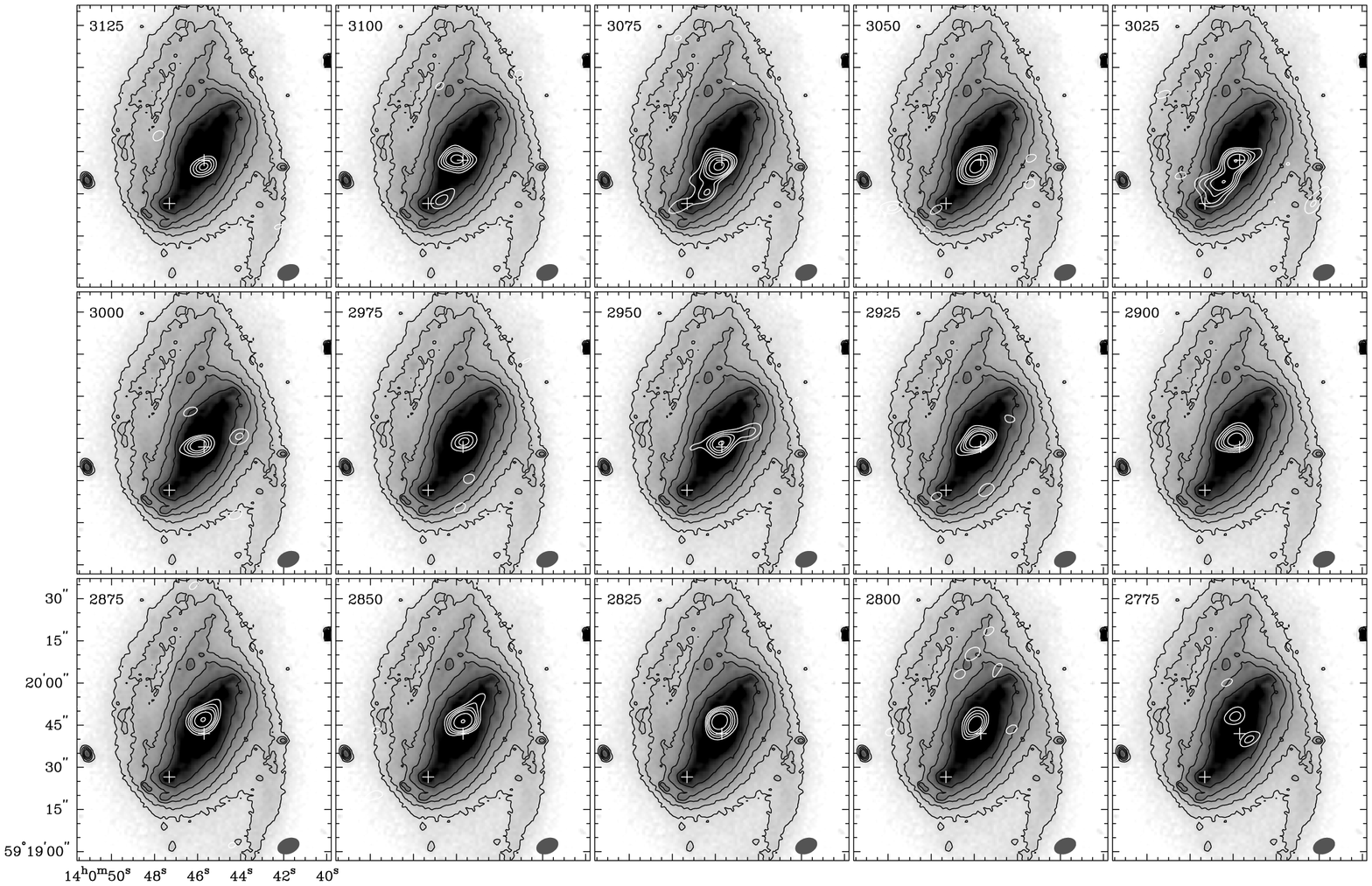} 
 \end{center}
\caption{Channel map of $^{13}$CO (1--0) in white contours overlaid on the SDSS $i$-band image in black contours and grey scale. The symbols are the same as in Figure \ref{fig_co_ch}. The contours are 2, 3, 4, 5, 7, and 10 $\sigma_{\mathrm{RMS}}$, where 1 $\sigma_{\mathrm{RMS}}$ is 6 mJy. Beam size of $^{13}$CO (1--0) is 8.1$\arcsec$ $\times$ 5.5$\arcsec$ (1.7 $\times $1.1 kpc) with P.A. $=$ -68.0$^{\circ}$. Note that the channel  velocity width of  $^{13}$CO (1--0) is larger than $^{12}$CO (1--0). Velocity resolution of $^{13}$CO (1--0) data cube is 25 km s$^{-1}$.
}
\label{fig_co13_ch}
\end{figure*}

Figure \ref{fig_co13_ch} shows the channel map of $^{13}$CO (1--0) (hereafter $^{13}$CO). The keys are the same as in Figure \ref{fig_co_ch}. $^{13}$CO emission was generated in the same velocity range as that of $^{12}$CO. The galactic center was solidly detected ($\sim$ 14 $\sigma_{\mathrm{RMS}}$ or 0.084 Jy beam$^{-1}$). The eastern bar was detected with the significance of $\sim$ 8 $\sigma_{\mathrm{RMS}}$ or 0.048 Jy beam$^{-1}$, while the other side was barely detected with $\sim$ 3 $\sigma_{\mathrm{RMS}}$ or 0.018 Jy beam$^{-1}$. The $^{13}$CO of the asymmetric arm was observed in the velocity  of $\sim$ 2800 km s$^{-1}$, consistent with the velocity of $^{12}$CO emission. We did not detect any $^{13}$CO from the two primary arms.

\subsubsection{Integrated Intensity Maps of CO}
\label{sec_int_co}

\begin{figure}[h]
\begin{center}
    \begin{minipage}{0.32\textwidth}
        \centering
		\includegraphics[width=1\textwidth, angle=0]{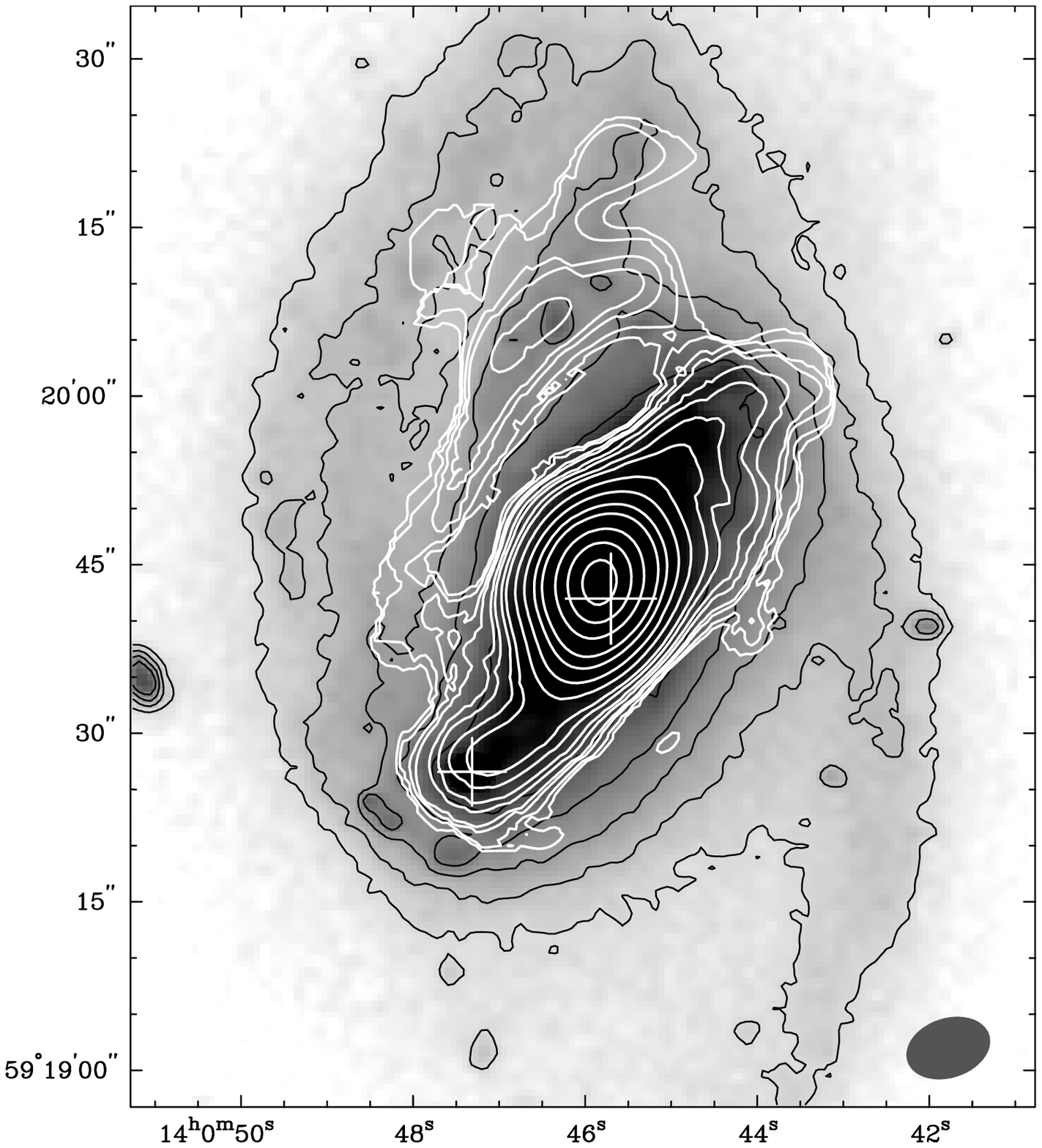} 
\begin{center}
 (a)
\end{center}
    \end{minipage}
    \begin{minipage}{0.32\textwidth}
        \centering
		\includegraphics[width=1\textwidth, angle=0]{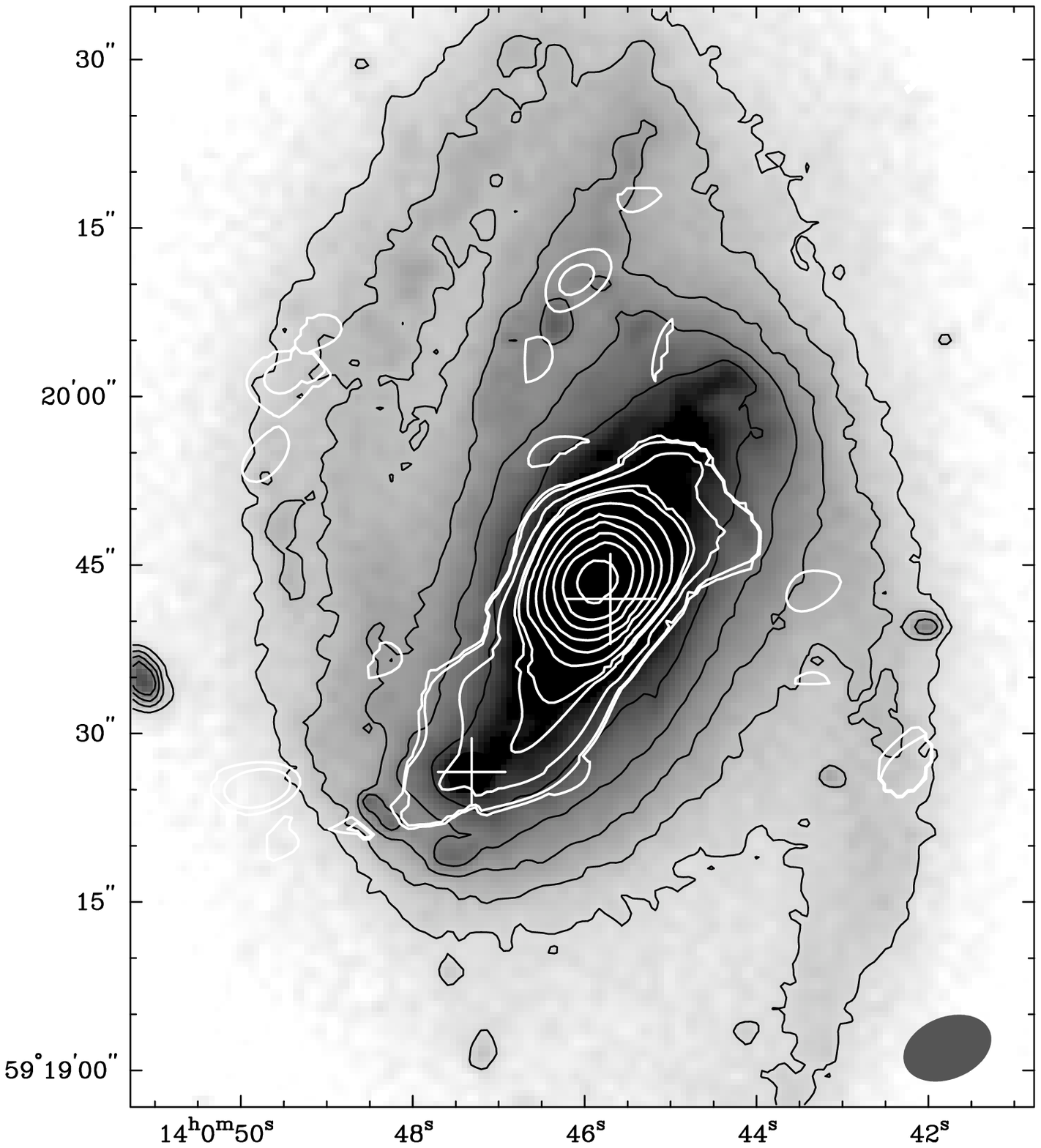} 
\begin{center}
 (b)
\end{center}
    \end{minipage}
    	\end{center}
	\caption{(a) Integrated intensity map of $^{12}$CO (1--0) (white contours) overlaid on SDSS $i$-band image (black contours and grey scale). Contour of $^{12}$CO (1--0) is plotted with steps of 0.6, 0.9, 1.5, 1.9, 3,  4, 6, 8, 13, 20, 30, 40, 60, 80, 100 Jy beam$^{-1}$ km s$^{-1}$. The galactic  center and the W-R region are marked with crosses. The noise level of the intensity map is $\sim$ 0.2 Jy beam$^{-1}$ km s$^{-1}$. (b)  Integrated intensity map of $^{13}$CO (1--0) (white contours) superimposed on SDSS $i$-band image. Contour of $^{13}$CO is displayed with steps of  0.3, 0.4, 0.9, 1.7, 2.0, 3.0, 4.5, 6.0, 8.0, 10, 13 Jy beam$^{-1}$ km s$^{-1}$. The noise level is $\sim$ 0.1 Jy beam$^{-1}$ km s$^{-1}$.}
	\label{fig_mom0}
\end{figure}

Integrated intensity maps of $^{12}$CO and $^{13}$CO are shown in Figure \ref{fig_mom0}(a) and \ref{fig_mom0}(b)  with white contours, overlaid on the SDSS $i$-band image. The morphology of molecular gas at the center and throughout the bar appears to be not severely disturbed by the galactic mergers. Both $^{12}$CO and $^{13}$CO intensity maps contain a bright central component with the radius of $\sim$ 1.5 kpc. There is an offset between the strongest CO intensity and the galactic center. The reason for this is enclosure of the galactic nucleus within a ring-like structure \citep{Gar96}; thus, the CO peak intensity position is determined from the convolution of the ring with our CO beam. Therefore, the CO peak is not necessarily at the galactic center if the gas distribution in the ring is not uniform. Offset ridges at the two sides of the bar are seen in $^{12}$CO, and are symmetric with respect to the galactic center. The offset ridges extend from the center to the distance of $\sim$ 5.6 kpc, and have the width of $\sim$ 1.2 kpc. The eastern ridge exhibits stronger $^{12}$CO intensity than the western ridge. Within this resolution, it cannot be determined whether the eastern ridge is, in general, stronger throughout the entire ridge or whether the W-R region, containing extra molecular gas contributed by the satellite, has been convolved with a large beam. $^{13}$CO is only detected in the eastern ridge. The length and width are consistent with those observed for $^{12}$CO. Because the two sides of the bar exhibit different star formation activity (stronger at the eastern ridge and weaker at the western ridge), their intrinsic $^{12}$C, $^{13}$C, $^{12}$CO, and $^{13}$CO abundances can be different. Both $^{12}$C and $^{13}$C can be enriched by the rapid cycling of interstellar medium (ISM) through stars. Once they are used to form $^{12}$CO and $^{13}$CO, the molecules can be destroyed by photo-dissociation through the UV photons from stars. The intrinsic $^{12}$CO/$^{13}$CO abundance ratio then depends on the net result of these processes (e.g., \cite{Mil05,Hen14,Szh14} and references therein). The observed line intensity ratio of $^{12}$CO/$^{13}$CO also depends on the radiative transfer and on the excitation of molecular lines. Therefore, star formation activity and the ISM environment play critical roles in determining the observed $^{12}$CO/$^{13}$CO ratio. The undetected $^{13}$CO of the western ridge may imply non-similar physical properties of molecular gas across the two sides of the bar.


The asymmetric spiral arm is connected to the eastern ridge in $^{12}$CO.
$^{12}$CO emission of  the asymmetric spiral arm is spatially  correlated with  the stellar light traced in SDSS image.
The asymmetric spiral arm  is more fragmented in  $^{13}$CO, and is  not connected to the eastern ridge. The detected $^{13}$CO is found in the regions with high $^{12}$CO intensity  and stellar light.

\subsubsection{Central Concentration of Molecular Gas and Bar Strength}
The total gas mass (H$_{2}$ $+$ He) in the central $\sim$ 1 kpc of NGC 5430 is considerably higher than other strong bar (SB and SAB) galaxies. We estimated the mass of gas in the central $\sim$ 1 kpc (in diameter) using the flux of the beam ($\sim$ 110 Jy beam$^{-1}$ km s$^{-1}$) at the galactic center, along with the  Galactic CO-to-H$_{2}$ conversion factor ($X_{\mathrm{CO}}$) of 2 $\times$ 10$^{20}$ cm$^{-2}$ (K km s$^{-1}$)$^{-1}$ \citep{Bol13}, and a correction factor of 1.36 for He and heavy elements.
Here we assumed that $X_{\mathrm{CO}}$ is constant and similar to that of the Milky Way disk across entire NGC 5430 when the gas properties are averaged over several 100 parsec to kiloparsec scale (e.g., \cite{Don13,Bol13}). 
The derived total mass is  $\sim$ (8.3 $\pm$ 1.5) $\times$ 10$^{8}$ M$_{\solar}$\footnote{The uncertainty is dominated by the uncertainty of flux calibration of CARMA observations.}. The value is considerably higher than the average value of galaxies with the same morphological type, $\sim$ 3.3 $\times$ 10$^{8}$ M$_{\solar}$ \citep{Sak99}\footnote{Corrected for the adopted $X_{\mathrm{CO}}$ in this work.}.

\citet{Kun07} show that the central concentration of molecular gas is proportional to the bar strength because strong bar exerts large gravitational torque and be most effective in driving gas inflows (see their Figure 49). Central concentration of molecular gas ($f_{\mathrm{con}}$) is defined as:
\begin{equation}
f_{\mathrm{con}}=\frac{M_{\mathrm{H_{2}}}(R_{\mathrm{K20}}/8)}{M_{\mathrm{H_{2}}}(R_{\mathrm{K20}}/2)},
\label{eq_fcon}
\end{equation}
where $R_{\mathrm{K20}}$ is the radius of 20 mag/arcsec$^{-2}$ in the 2MASS $K_{\mathrm{S}}$-band, and $M_{\mathrm{H_{2}}}(R_{\mathrm{K20}}/8)$ and $M_{\mathrm{H_{2}}}(R_{\mathrm{K20}}/2)$ represent the gas mass enclosed within the radius of  $R_{\mathrm{K20}}$/8 and $R_{\mathrm{K20}}$/2, respectively. Using Equation (\ref{eq_fcon}), the $f_{\mathrm{con}}$ of NGC 5430 is estimated as $\sim$ 0.34. The value of $f_{\mathrm{con}}$ is indeed higher that the average value (0.20) of barred galaxies in \citet{Kun07}\footnote{Galaxies in \citet{Kun07} have distance of $<$ 25 Mpc and an observing resolution of 15$\arcsec$. NGC 5430 is farther than those galaxies by $\sim$ 2 times, but observed with a resolution $\sim$ 2 times finer. Thus these two studies sample the same galactocentric ranges in Equation (\ref{eq_fcon}).}. Therefore, NGC 5430 likely has a stronger bar than the visually isolated galaxies in \citet{Kun07} in terms of $f_{\mathrm{con}}$.

The strong bar is also implied by the high de-projected bar ellipticity. There is increasing evidence that bar morphology is determined primarily by the bar strength. Various studies, including both observations and simulations, have reported the existence of correlation between the bar strength and bar ellipticity  \citep{Ath92,Mar97,Abr00,Blo01,Blo04,Com10,Kim12}. The Spitzer project of S$^{4}$G \citep{She10} has decomposed the galactic structure of NGC 5430 by using the GALFIT package and the Spitzer IRAC 3.6 $\mu$m image, by assuming a three-component bulge-disk-bar model  \citep{Sal15}. The derived bar ellipticity is ~0.6. The bar ellipticity of NGC 5430 is again on the higher side of the observed and simulated bar ellipticity, which ranges from ~0.2 to 0.8.

It is certainly possible that the formation of the strong bar in NGC 5430 is induced by the past galaxy interaction. Simulations have shown that even a small perturbation, such as galaxy flyby and minor merger, can induce a bar (e.g., \cite{Nog87,Ber04,Lang14,Lok14}).

\subsection{Velocity Fields}
\label{sec_kin}
\subsubsection{Velocity Fields of CO and the Comparison with H$\alpha$}

\begin{figure*}[ht]
\begin{center}
    \begin{minipage}{0.19\textwidth}
        \centering
		\includegraphics[width=1\textwidth, angle=0]{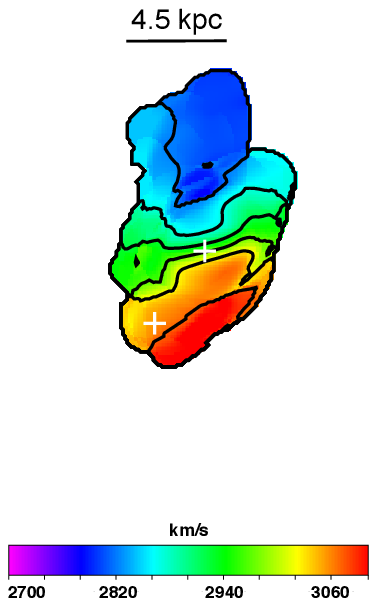} 
\begin{center}
 (a)
\end{center}
    \end{minipage}
    \begin{minipage}{0.19\textwidth}
        \centering
		\includegraphics[width=1\textwidth, angle=0]{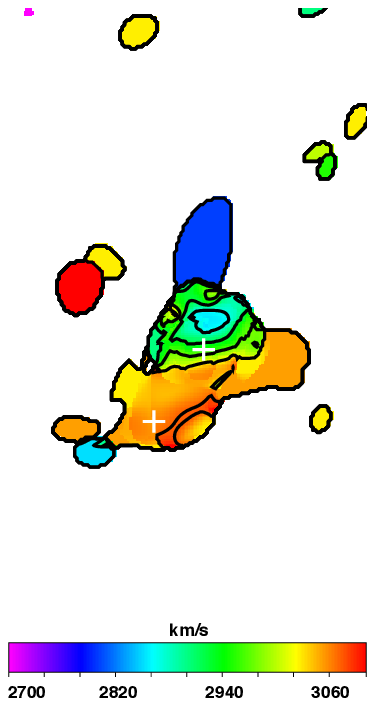} 
\begin{center}
 (b)
\end{center}
\end{minipage}
    \begin{minipage}{0.19\textwidth}
        \centering
		\includegraphics[width=1\textwidth, angle=0]{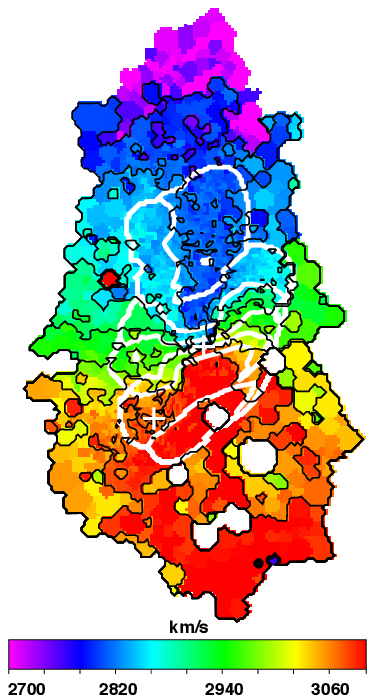} 
\begin{center}
 (c)
\end{center}
    \end{minipage}
    \begin{minipage}{0.19\textwidth}
        \centering
		\includegraphics[width=1\textwidth, angle=0]{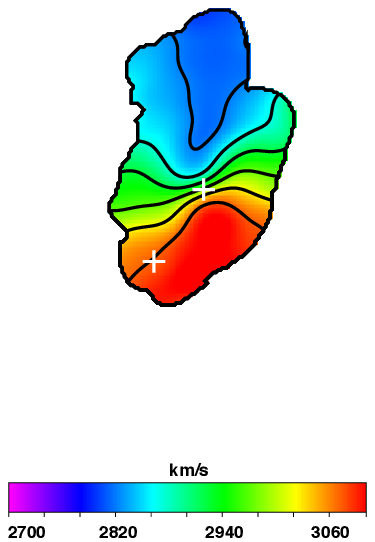} 
\begin{center}
 (d)
\end{center} 
\end{minipage}   
\begin{minipage}{0.19\textwidth}
        \centering
		\includegraphics[width=1\textwidth, angle=0]{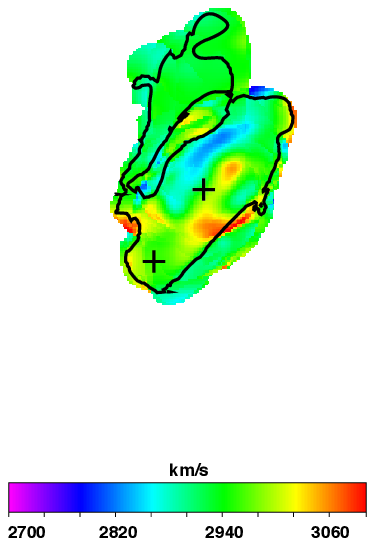} 
\begin{center}
 (e)
\end{center}
    \end{minipage}
    	\end{center}
	\caption{Velocity fields of NGC 5430. Color scale of  velocity fields is shown with a range 2700 to 3100 km s$^{-1}$. Contours of velocity fields are plotted in steps of 2720, 2770, 2820,  2870, 2920, 2970, 3020, and 3070 km s$^{-1}$ in all panels.  The galactic center and the W-R region are marked with crosses in all panels as well. (a) Velocity field of $^{12}$CO (1--0) (color scale and contours). (b) Velocity field of $^{13}$CO (1--0) (color scale and contours). (c) Comparison of velocity fields of H$\alpha$ (color scale and black contours) and $^{12}$CO (white contours). (d) Smoothed velocity field of H$\alpha$ with velocity and spatial resolutions of $^{12}$CO field. (e) Residual velocity of the smoothed velocity field of H$\alpha$ (panel d) minus the velocity field of $^{12}$CO field (panel a).}
	\label{fig_mom1}
\end{figure*}

Figures \ref{fig_mom1}(a) and \ref{fig_mom1}(b) show the velocity fields of $^{12}$CO and $^{13}$CO, respectively. Two crosses mark the galactic center (upper cross) and the W-R region (lower cross). The receding and approaching sides are located in the south and north, respectively. The velocity fields of $^{12}$CO and $^{13}$CO are consistent with each other on the global scale. 

The detailed velocity distribution in the $^{13}$CO field is affected by the low sensitivity. For this reason, the kinematic discussion will focus on the $^{12}$CO field. The central region reveals parallel velocity contours in $^{12}$CO. However, velocity fields of disk galaxies usually exhibit either the ``spider''  diagram of circular motion or the ``S-shape''  twisted velocity gradient of non-circular motion. Our CO emission may be smeared out owing to the large beam size and low velocity resolution (12 km s$^{-1}$) at the galactic center, where velocity field changes rapidly within a single beam. Therefore neither the spider nor the S-shape pattern is seen. It is marginally observed from the $^{12}$CO map that the velocity gradient is almost normal to the CO offset ridges (namely the contours are parallel to the ridges), implying a large velocity jump across the bar, which is predicted by theoretical studies (e.g., \cite{Ath92}). For $^{12}$CO, this velocity jump is about 100 km s$^{-1}$ per 1 kpc. 

Figure \ref{fig_mom1}(c) compares the velocity field of $^{12}$CO (white contours) to the H$\alpha$ from the GHASP survey (color scale and black contours). Spatial and velocity resolutions of the H$\alpha$ field are $\sim$ 4$\arcsec$ (820 pc) and 5 km s$^{-1}$, respectively. The CO field reveals the global variation in velocity that is observed in H$\alpha$. Because of the higher resolution, the velocity field of H$\alpha$ reveals more velocity structures at the galactic center. The galactic center reveals twisted S-shape velocity contours,  which is an evidence of the bar non-circular motion \citep{Epi08}. This pattern is barely seen if the velocity field of H$\alpha$ is smoothed to the   velocity and spatial resolutions of the CO maps (Figure \ref{fig_mom1}(d)). Figure \ref{fig_mom1}(e) shows the velocity residual (color scale) of the smoothed H$\alpha$ field with the CO field subtracted from it, overlaid on the emission boundary of the $^{12}$CO integrated intensity map. The majority of the residuals are within $\pm$ 10 km s$^{-1}$, while some regions deviate by as much as $\pm$ 50 km s$^{-1}$. The regions with large residuals are located around the edge of the CO emissions or are associated with weak CO emission. Thus, these large residuals are not significant for judging the difference between molecular and ionized gases. 

\begin{figure}
\begin{center}

    \begin{minipage}{0.15\textwidth}
        \centering
		\includegraphics[width=1\textwidth, angle=0]{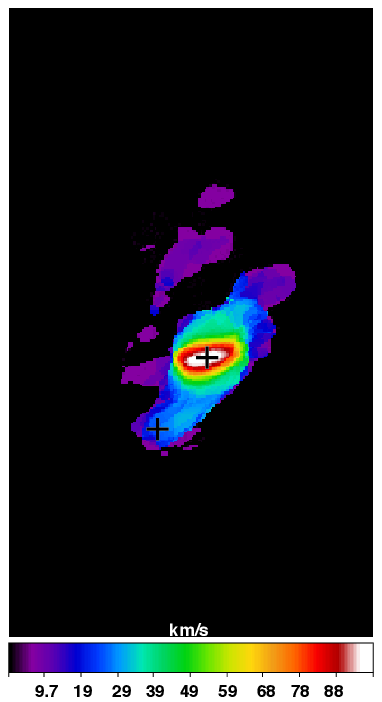} 
\begin{center}
 (a)
\end{center}
    \end{minipage}
    \begin{minipage}{0.15\textwidth}
        \centering
		\includegraphics[width=1\textwidth, angle=0]{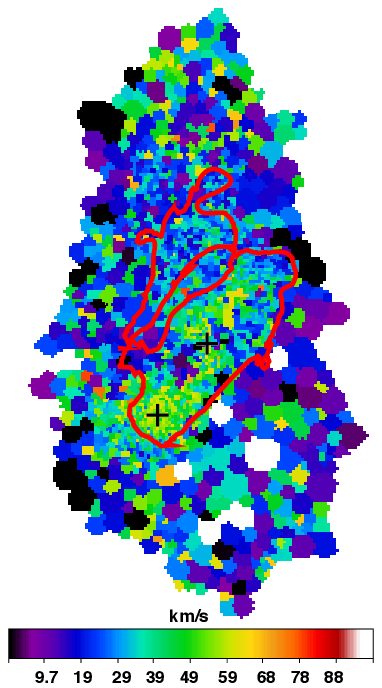} 
\begin{center}
 (b)
\end{center}
    \end{minipage}
        \begin{minipage}{0.15\textwidth}
        \centering
		\includegraphics[width=1\textwidth, angle=0]{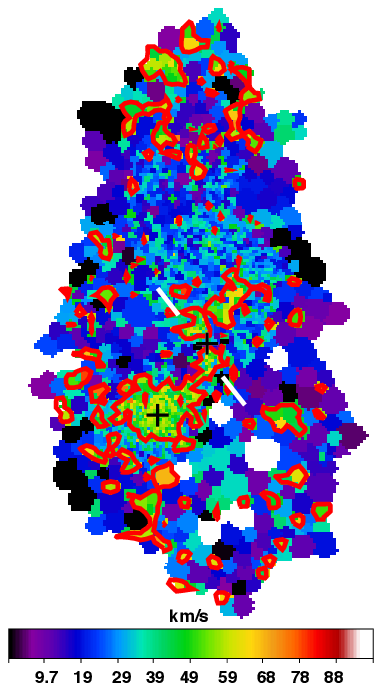} 
\begin{center}
 (c)
\end{center}
    \end{minipage}
    	\end{center}
	\caption{Velocity dispersions of NGC 5430. The galactic center and the W-R region are marked with black crosses in all panels (a) Velocity dispersion (second moment) of $^{12}$CO (1--0). (b) Velocity dispersion of H$\alpha$ (color scale) with the boundary of $^{12}$CO integrated intensity map overlaid (red contour). (c) Velocity dispersion of H$\alpha$ (color scale) with the H$\alpha$ velocity dispersion of 38 km s$^{-1}$ overlaid (red contour). Two white lines mark the locations of the intersection between  $x_{1}$ and $x_{2}$ orbits (contact points).}
	\label{fig_veldisp}
\end{figure}

Velocity dispersions of molecular and ionized gases appear to be more different. Figure \ref{fig_veldisp}(a) shows the velocity dispersion of the molecular gas, while panels (b) and (c) show the velocity dispersion of the ionized gas in color scales. Velocity dispersion of the molecular gas reaches the peak of $\sim$ 90 km s$^{-1}$  at the galactic center, with an average of $\sim$ 65 km s$^{-1}$ within the central 1 kpc. In the bar region, large velocity dispersion appears at the leading side of the bar. The eastern bar exhibits a slightly larger dispersion than the western bar, $\sim$ 15 -- 40 km s$^{-1}$ compared with $\sim$ 12 -- 30 km s$^{-1}$. The asymmetric spiral shows a low velocity dispersion of 5 -- 12 km s$^{-1}$. We found that this distribution of velocity dispersion is consistent with that observed in low-resolution $^{12}$CO and can be explained by secular origin. The high velocity dispersion ($>$ 50 km s$^{-1}$) at the galactic center results from the accumulation of molecular gas in the collision region where $x_{1}$ (bar) and $x_{2}$ (circumnuclear region) orbits intersect, and from the spatial superposition of circular and noncircular components  (e.g., \cite{Hut00,Sch00,Hsi11}). The intermediate velocity dispersion (a few tenths km s$^{-1}$) of the bar, peaking at the leading side on both sides of the bar, is owing to the crowded obits  (e.g., \cite{Ath92,Hut00}). The low velocity dispersion ($<$ 15 km s$^{-1}$) in the asymmetric spiral arm is consistent with that of the unresolved giant molecular clouds (associations) in the galactic disks (e.g., \cite{Tos07,Don13,Fae14}).

In contrast to the CO, velocity dispersion of the ionized gas exhibits stronger local variation because of the higher resolution. Figures \ref{fig_veldisp}(b) and \ref{fig_veldisp}(c) show the velocity dispersion of H$\alpha$ using color scale. The CO boundary is overlaid with red contour in Figures \ref{fig_veldisp}(b). Two regions with high velocity dispersion are observed in the central region, located at the opposite sides of the galactic center. These regions correspond to the intersection of $x_{1}$ and $x_{2}$ orbits, the so-called contact points (indicated by two white lines in Figure\ref{fig_veldisp}(c)). The velocity dispersion of H$\alpha$ peaks at the northern intersection with 72 km s$^{-1}$, while the maximal velocity dispersion at the southern intersection is 64 km s$^{-1}$. The intersections are connected by two high-dispersion dust lanes of the bar. The maximal velocity dispersion at the dust lanes reaches $\sim$ 65 km s$^{-1}$. Beyond the dust lanes, the W-R region exhibits larger velocity dispersion than the opposite region in the western bar. The peak velocity dispersion in the W-R region is 61 km s$^{-1}$ with an average of 45 km s$^{-1}$ within the surrounding 1 kpc region. The large velocity dispersion of H$\alpha$ compared with the CO is perhaps owing to the bias towards the diffuse ionized gas disturbed by supernova explosions and the expanding nebulae around the massive populations, such as the W-R stars \citep{Mel99,Moi12,Kam15}.

\subsubsection{Position-Velocity Diagrams and Rotation Curve}
\label{sec_pv_rc}

\begin{figure}
 \begin{center}
  \includegraphics[scale=0.37, angle =90]{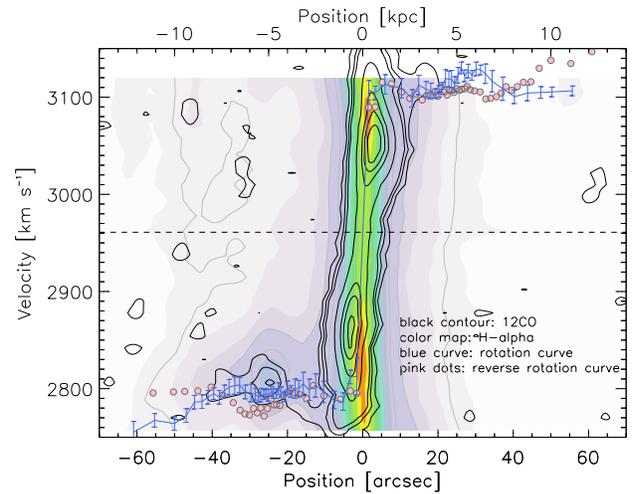} 
 \end{center}
\caption{Position-Velocity (PV) diagram of NGC 5430. PV diagram is cut along the major axis of NGC 5430 with an position angle of 181.9$^{\circ}$. PV diagram of $^{12}$CO (1--0) is presented with black contours. PV diagram of H$\alpha$ is shown with color scale and grey contours. The PV diagram of H$\alpha$  has the same spatial and velocity resolution with that of $^{12}$CO (1--0).  Blue curve denotes the  rotation curve traced by H$\alpha$ created by \citet{Epi08}. For comparing between the two sides of the galaxies, reserved rotation curve is overlaid with pink dots. 
}
\label{fig_pv}
\end{figure}

Figure \ref{fig_pv} compares the position-velocity (PV) diagrams of H$\alpha$ (color scales, grey contours), and $^{12}$CO (black contours). The PV diagrams are obtained along the major axis of the velocity fields at 181.9$^{\circ}$. The PV diagram of the ionized gas has the same resolution as that of the molecular gas. Both PV diagrams are composed of a circumnuclear disk with steep increase in velocity within $\pm$ 20$\arcsec$ ($\sim$ 4 kpc), and a strong emission at $\sim$ –30$\arcsec$ ($\sim$ 6 kpc) and 2800 km s$^{-1}$ tracing the asymmetric spiral arm. 

 The blue curve in Figure \ref{fig_pv} represents the rotation curve of the NGC 5430 measured in the ionized gas by \citet{Epi08}. For comparison between the receding and the approaching sides, the reversed rotation curve is overlaid with pink dots. The rotation curve reaches a plateau within a few arcsec on both sides, fluctuating at about 3100 km s$^{-1}$ ($+$ 140 km s$^{-1}$ relative to the $V_{sys}$ of 2960 km s$^{-1}$) and 2800 km s$^{-1}$ (- 160 km s$^{-1}$). The two sides of the rotation curve are symmetric out to $\sim$ $\pm$ 20$^{\arcsec}$ ($\sim$ 4 kpc). This radius includes the central region and majority of the bar. A small bump is observed at the receding side at the radius just beyond the W-R region, $\sim$ 30$\arcsec$ ($\sim$ 6 kpc), where the velocity at the receding side is $\sim$ 10 to 20 km s$^{-1}$ larger than that at the approaching side. The kinematic asymmetry around the W-R region might be owing to the inhomogeneous mass distribution caused by extra gas contributed by the satellite collisions and the subsequent active star formation (note that the rotation curve is derived from H$\alpha$ and therefore it is sensitive to the star formation history). 

In terms of the rotation curve, the NGC 5430 has almost relaxed and shows nearly no signs of interaction in the inner disk.
 Overall, we found that the rotation curve beyond the central region ($>$ $\left | \pm 1\, \mathrm{kpc} \right |$)  is flat, fluctuating within $\sim$  60 kms$^{-1}$ on the both sides. Moreover, the extent of asymmetry between the two sides is small ($<$ 50 km s$^{-1}$). Simulations reveal that systems with ongoing minor galaxy interaction or encounters have non-flat rotation curves. Velocity increases with galactocentric radius by at least 100 km s$^{-1}$ in the galactic disk. Moreover, the rotation curves are asymmetric, differing by $>$ 100 km s$^{-1}$ between the two sides of the disk \citep{Kro06}. Therefore, the observed rotation curve of NGC 5430 implies that the system has almost relaxed, exhibiting no obvious signs of kinematic distortion.

The outer disk shows a larger degree of asymmetry than the inner disk. Beyond the asymmetric spiral arm ($\pm$ 40$\arcsec$ or $\sim$ 8 kpc), the two sides of the rotation curve exhibit a larger deviation by $\sim$  30 km s$^{-1}$. It is possible that the outer disk is not completely stabilized from the past interaction with HOLM569B. Moreover, because NGC 5430 is the member of a galaxy group \citep{Gel83}, the perturbed outer disk can also result from ongoing interaction with other group members.

The kinematics of NGC 5430 helps to constrain the timescale on which the two galaxies encounter. Both simulations and observations suggest that the global kinematic disturbances of minor mergers fade within $\sim$ 1 -- 2 Gyr after the first encounter of two galaxies, and the rotation curves exhibit no more severe distortions  (e.g. \cite{Rub99,Dal01,Kro06}). Thus, NGC 5430 likely captured HOLM 569B around a few Gyr ago. However, the merger system is not fully settled yet as suggested by the large velocity dispersion of the ionized gas, inhomogeneous mass distribution at the collision site (the W-R region), and probably the asymmetric rotation curve at the outer disk as well.

\section{Physical Properties of Molecular Gas and Star Formation activity}
\label{sec_phy_prop_sf}
\subsection{Physical Properties of Molecular Gas}
\subsubsection{Line Ratios}
\label{sec_ratios}
Because of the different excitation conditions associated with each molecular line, the line ratio constrains the physical properties of the molecular gas. In addition to the $^{12}$CO (1--0) to $^{13}$CO (1--0) ratio ($R_{\mathrm{1-0}}$),  $^{12}$CO (2--1) to $^{12}$CO (1--0) ($R_{\mathrm{12CO}}$) and $^{12}$CO (2--1) to $^{13}$CO (2--1)  ($R_{\mathrm{2-1}}$) ratios are quoted by using the measurements reported in the literature. Table \ref{TAB_line_ratios}  shows the line intensities and ratios of the galactic center and the W-R region of NGC 5430. All values have been measured on a scale of 22$\arcsec$ (in diameter), centering on the galactic center and the W-R region. The scale was chosen by the lowest resolution among all measurements, which is the IRAM 30m single dish observation in $^{13}$CO (2--1).  

The line ratios of the galactic center of NGC 5430 are similar to those of the   Galactic star forming regions,  starburst nucleus, and unresolved starburst galaxies, while the values for the W-R region are consistent with the galactic disk clouds. $R_{\mathrm{1-0}}$ of the galactic center and the W-R region of NGC 5430 are $\sim$ 10 and $\sim$ 6, respectively. The high $R_{\mathrm{1-0}}$ (10 -- 20) is observed in the  Galactic star forming regions,  starburst nucleus, and unresolved starburst galaxies, while the low $R_{\mathrm{1-0}}$ ($\sim$ 6) occurs in the galactic disk, relatively quiescent molecular clouds (e.g., \cite{Aal95,Pag01,Sim01,Tan11,Pap12}). The high $R_{\mathrm{1-0}}$ is explained by the change in opacity of the molecular gas owing to the high temperature,  velocity dispersion and column density of  active star forming region and galaxies. $R_{\mathrm{12CO}}$ also increases with increasing star formation activity. The variation stems from the fact that the upper energy boundary ($\sim$ 16.6 K) and the effective (including radiative trapping) critical density ($\sim$ 230 cm$^{-3}$) of $^{12}$CO (2--1) are slightly higher than the typical temperature ($\sim$ 10 K) anddensity ($\sim$100 cm$^{-3}$) of molecular clouds. Therefore, $R_{\mathrm{12CO}}$ increases from $\sim$ 0.5 in quiescent molecular clouds, to $>$ 0.7 in dense and/or warm star forming clouds (e.g., \cite{Chi94,Oka98,Saw01,Kod12}). The line ratio for the W-R region (0.54) is similar to those of the quiescent clouds, while a higher ratio (0.7), close to that of star forming regions and galaxies, is found for the central region. Finally, $R_{\mathrm{2-1}}$ appears to increase from $<$ 10 to 10 -- 60 when moving from  quiescent to  star forming regions and galaxies  (e.g., \cite{Aal95,Saw01,Pap12,Isr15}). The trend is similar to that of $R_{\mathrm{1-0}}$. $R_{\mathrm{2-1}}$ is available at the central region of NGC 5430. The measured value of 12.6 was in the range for  star forming regions and galaxies.

\subsubsection{Large Velocity Gradient Calculations}
\label{sec_lvg}
Gas temperature and volume density can be quantified by comparing the observed line ratios with those obtained by using one-zone large velocity gradient (LVG) models based on local photon trapping and the escape probability method \citep{Gol74,Sol74}. In the LVG calculations, molecular excitation conditions depend on the local kinetic temperature ($T_{\mathrm{k}}$), line opacity (column density of CO per unit velocity, $N_{\mathrm{CO}}$/$dv$), and volume density ($n_{\mathrm{H_{2}}}$), as low-$J$ CO is excited by the collisions with H$_{2}$. The CO-H$_{2}$ collisional cross-sections reported by  \citet{Yan10} were used in the present work. 

The column density of CO per unit velocity was assumed by using our observations, and it is $\log( N\mathrm{_{CO}}/dv)\,[\mathrm{cm^{-2}(K\,km\,s^{-1})^{-1}}]=16.6\,-\,17.2$. Thus, the calculation was simplified to consider the  variables of $T_{\mathrm{k}}$ and $n_{\mathrm{H_{2}}}$. Assuming the width of velocity range for giant molecular clouds associations to be 40 -- 100 km s$^{-1}$, and assuming the surface density of $\sim$ 1000  M$_{\solar}$ pc$^{-2}$ and a CO-to-H$_{2}$ abundance of 8 $\times$ 10$^{-5}$, the resultant $\log( N\mathrm{_{CO}}/dv)$ is in the above range. We note that the following results of our LVG calculations were based on the above assumptions, and therefore should be treated with caution. 

\begin{figure}[h]
    \begin{minipage}{0.3\textwidth}
        \centering
		\includegraphics[width=1\textwidth, angle=90]{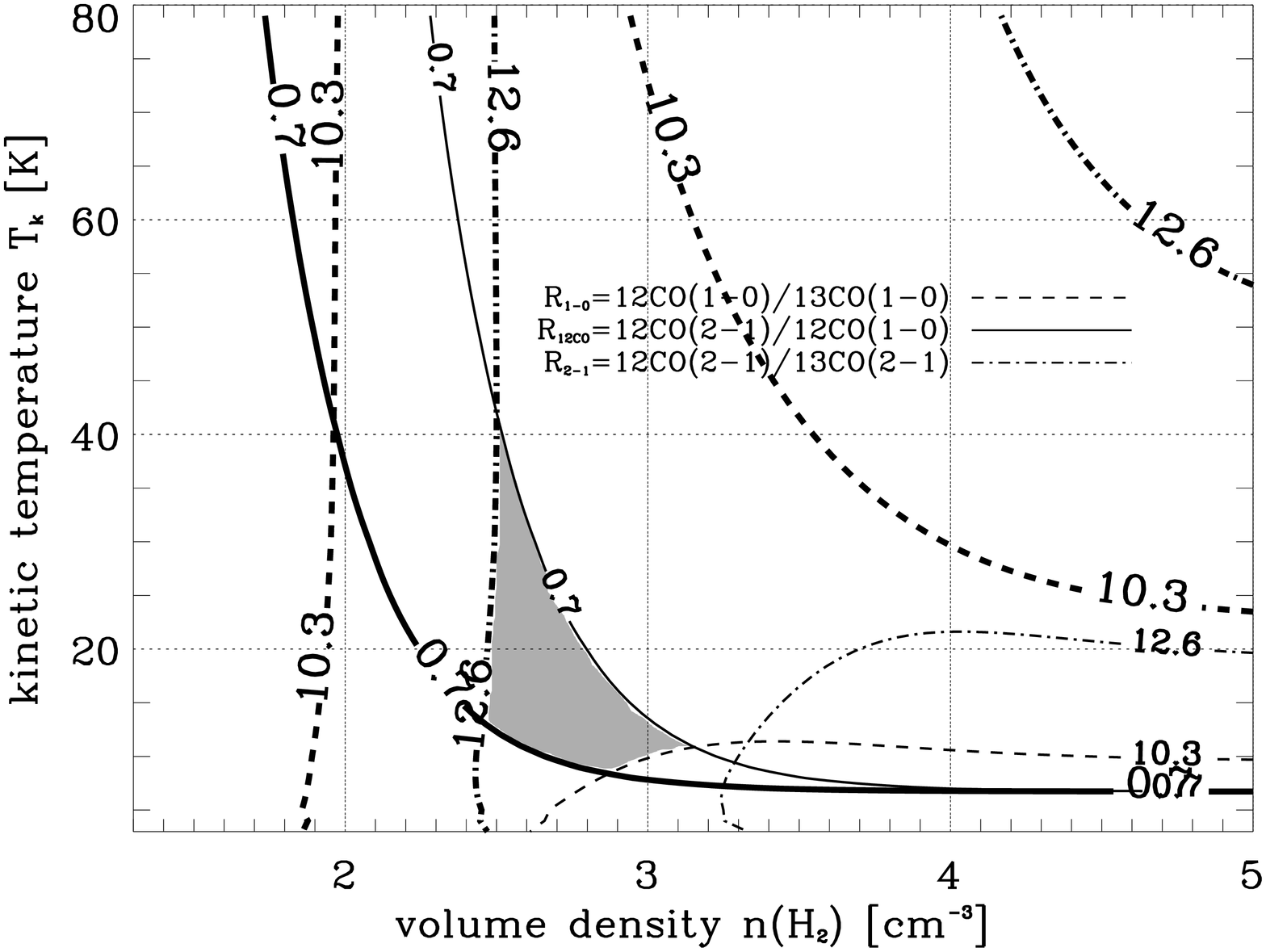} 
\begin{center}
 (a)
\end{center}
    \end{minipage}
    \begin{minipage}{0.3\textwidth}
        \centering
		\includegraphics[width=1\textwidth, angle=90]{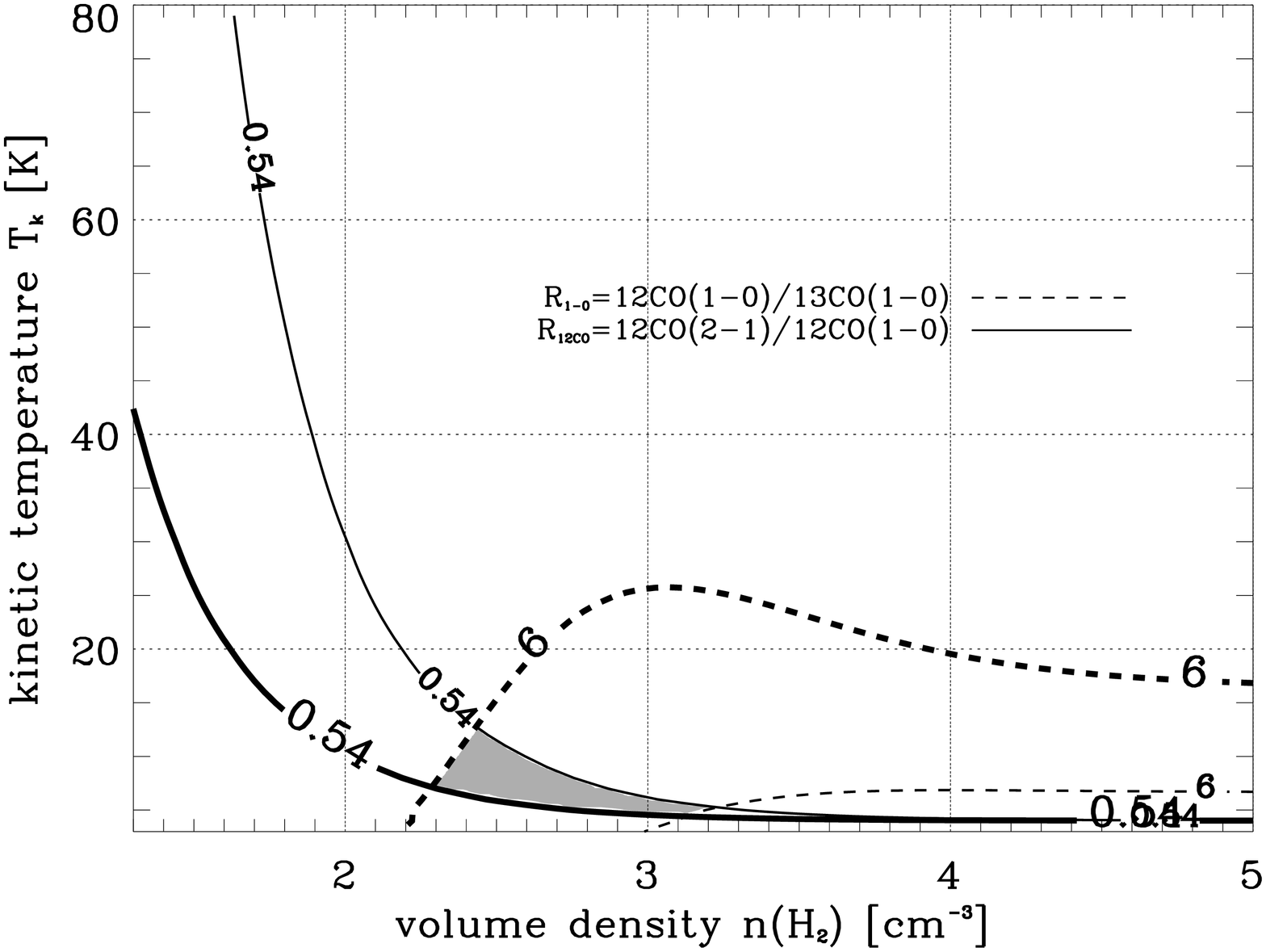} 
\begin{center}
 (b)
\end{center}
    \end{minipage}
	\caption{Result of LVG calculations in (a) the central region and (b) the W-R region. The dashed, solid and dotted-dashed contours represent the ratios of $^{12}$CO(1--0)/$^{13}$CO(1--0) ($R_{\mathrm{1-0}}$), $^{12}$CO(2--1)/$^{12}$CO(1--0) ($R_{\mathrm{12CO}}$), and $^{12}$CO(2--1)/$^{13}$CO(2--1) ($R_{\mathrm{2-1}}$), respectively. Ratio of $R_{\mathrm{2-1}}$ is available for the central region only.  For each ratio, two contours are drew, representing the $\log( N\mathrm{_{CO}}/dv)$ $=$ 16.6 (thin contour) and 17.2 (thick contour) $\mathrm{cm^{-2}(K\,km\,s^{-1})^{-1}}$, respectively. Solutions of LVG calculations are highlighted with grey shadows. }
	\label{fig_lvg}
\end{figure}

Figure \ref{fig_lvg} shows the results of our LVG calculations. Panel (a) and (b) show the results for the central and W-R regions, respectively. The results are shown as $T_{\mathrm{k}}$ (y-axis) versus $n_{\mathrm{H_{2}}}$ (x-axis). The observed line ratios are shown as contours. The dash, solid and dotted-dashed contours denote $R_{\mathrm{1-0}}$, $R_{\mathrm{12CO}}$, and $R_{\mathrm{2-1}}$, respectively.  $R_{\mathrm{2-1}}$ is available for the central region only. For each line ratio, two contours are plotted, corresponding to $\log( N\mathrm{_{CO}}/dv)$ $=$ 16.6 in thin contour and 17.2 in thick contour. Solutions were found in the areas enclosed by all contours; these areas are marked in grey. 

The LVG calculations suggest that the bulk molecular gas is not heated by the compact starburst in the W-R region, whereas the bulk molecular gas is generally warm in the central region. Bulk molecular gas associated with the W-R region has low temperature of $\sim$ 10 K, similar to that of the galactic disk molecular gas. Starburst is ongoing in the W-R region (\S\ref{sec_sfr}); thus, the temperature is expected to be high, presumably at or above $\sim$ 100 K  (e.g., \cite{Wei01,Mat10}). The low calculated temperature implies that the off-center starburst is spatially small so that the bulk molecular gas seen in low-resolution observations (average over the scale of kpc) remains cold. The $T_{\mathrm{k}}$ in the central region is 10 -- 40 K, warmer than the galactic disk molecular gas ($\sim$ 10 K). Although the star forming activity in the central region is not as drastic as that in the W-R region, the distribution of star forming regions is more extended. Thus, the molecular gas is generally warm in the central region,  yielding a high average $T_{\mathrm{k}}$  over the scale of kpc.

The LVG calculations show that the density of bulk molecular gas traced by the CO lines is 200 --1600  cm$^{-3}$ in both regions. The gas density is consistent with the effective critical density of the CO lines.

\begin{table*}
\caption{Line intensities and ratios of the central (second row) and W-R   regions (third row). Measurements of $^{12}$CO and $^{13}$CO (1--0) are from this work using CARMA telescope. $^{12}$CO (2--1) and HCN (1--0) are obtained from the IRAM 30-m telescope by \citet{Con97}. $^{13}$CO (2--1)  is measured using IRAM 30-m telescope as well by \citet{Kru90}. The listed numbers have a resolution of 22$\arcsec$ in diameter. The original resolution of HCN (1--0) is 27$\arcsec$. The total flux (in Jansky) within the central 27$\arcsec$ has been converted into the brightness of a 22$\arcsec$ beam based on the assumption  that the spatial size of dense gas cannot be larger than that of CO emission.}
\label{TAB_line_ratios}
\begin{tiny}
\centering
\begin{tabular}{cccccccccc}
\hline
$^{12}$CO (1-0) & $^{13}$CO (1-0) & $^{12}$CO (2-1) & $^{13}$CO (2-1)  & HCN (1-0)       & \multirow{2}{*}{$R_{\mathrm{1-0}}$$=$$\frac{^{12}\mathrm{CO\, (1-0)}}{\mathrm{^{13}CO\, (1-0)}}$} & \multirow{2}{*}{$R_{\mathrm{12CO}}$$=$$\frac{^{12}\mathrm{CO\, (2-1)}}{\mathrm{^{12}CO\, (1-0)}}$}& \multirow{2}{*}{$R_{\mathrm{2-1}}$$=$$\frac{^{12}\mathrm{CO\, (2-1)}}{\mathrm{^{13}CO\, (2-1)}}$}   & \multirow{2}{*}{$R_{\mathrm{dense}}$$=$$\frac{\mathrm{HCN\, (1-0)}}{\mathrm{^{12}CO\, (1-0)}}$}  \\

 [K km s$^{-1}$] & [K km s$^{-1}$] & [K km s$^{-1}$] & [K km s$^{-1}$] & [K km s$^{-1}$] &                                                  &       &                                                                                       \\
\hline
117.4 $\pm$ 2.1 & 12.7 $\pm$ 0.1  & 81.6 $\pm$ 0.5 & 6.5  & 2.1 $\pm$ 0.3  & 10.3 $\pm$ 0.1     & 0.70 $\pm$ 0.01   &12.6    &  0.018 $\pm$ 0.001          \\
 11.4  $\pm$ 0.3  & 1.9 $\pm$ 0.1   & 6.1 $\pm$ 0.5& $\dots$  & $<$ 0.8    & 6.0 $\pm$ 0.1              & 0.54 $\pm$ 0.01   & $\dots$ &  $<$ 0.070    \\            
   \hline               
\end{tabular}
\end{tiny}
\end{table*}

\subsection{Star Formation Activity}
\label{sec_sfr}

\subsubsection{Star Formation Rate}
\label{sec_global_sfr}
\begin{figure*}[ht]
\begin{center}

    \begin{minipage}{0.2\textwidth}
        \centering
		\includegraphics[width=1\textwidth, angle=0]{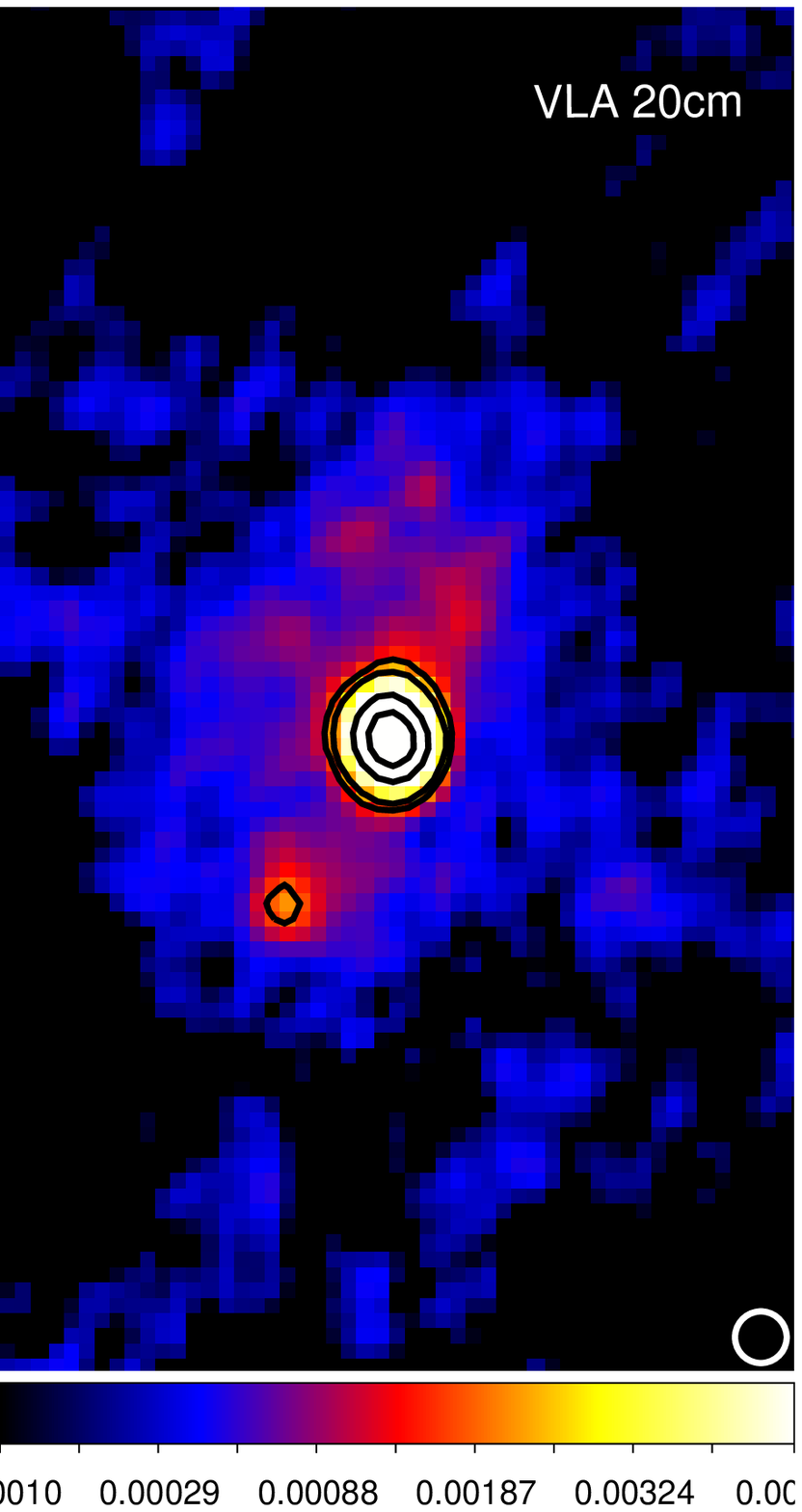} 
\begin{center}
 (a)
\end{center}
    \end{minipage}
    \begin{minipage}{0.2\textwidth}
        \centering
		\includegraphics[width=1\textwidth, angle=0]{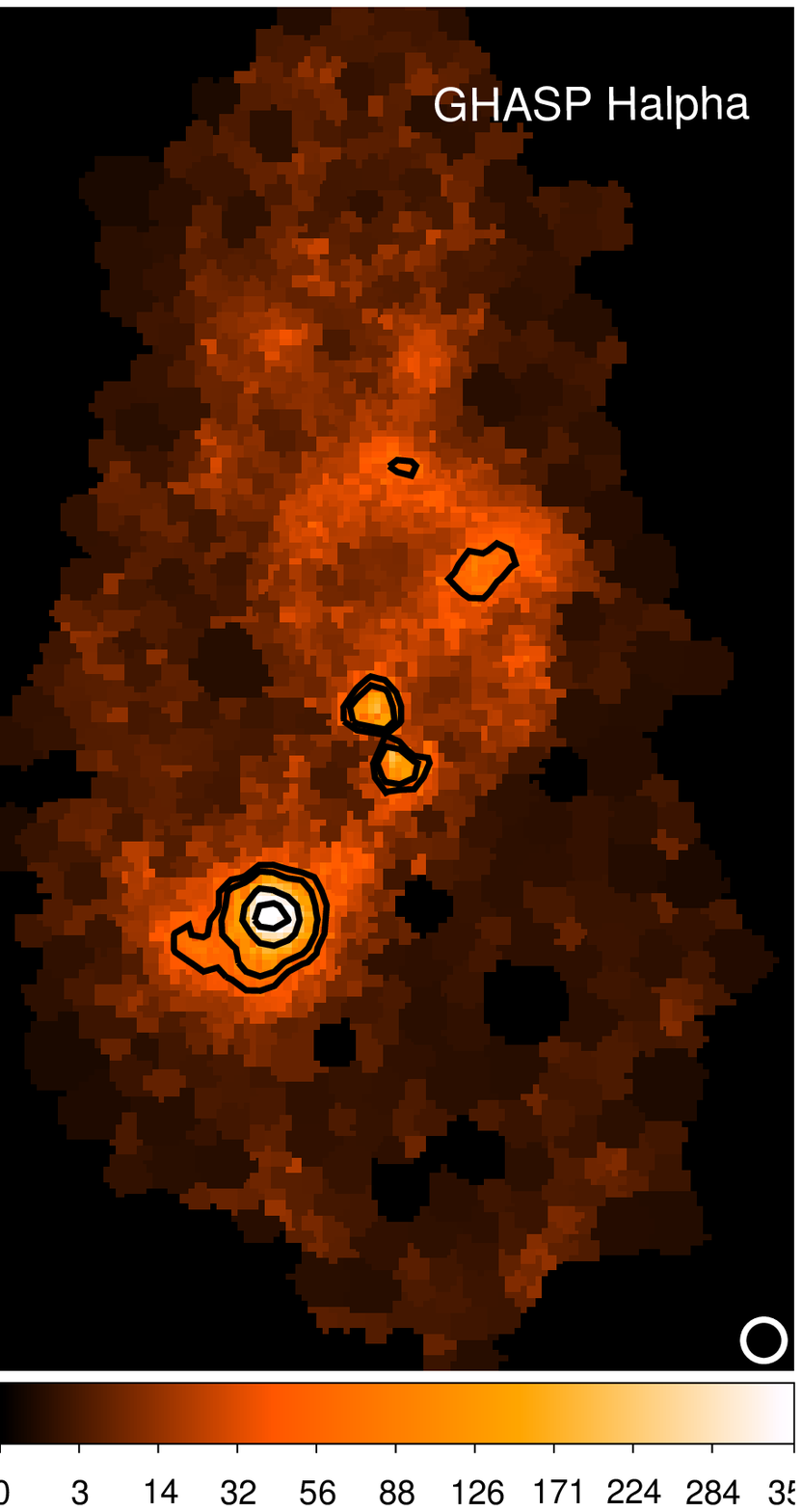} 
\begin{center}
 (b)
\end{center}
    \end{minipage}
        \begin{minipage}{0.2\textwidth}
        \centering
		\includegraphics[width=1\textwidth, angle=0]{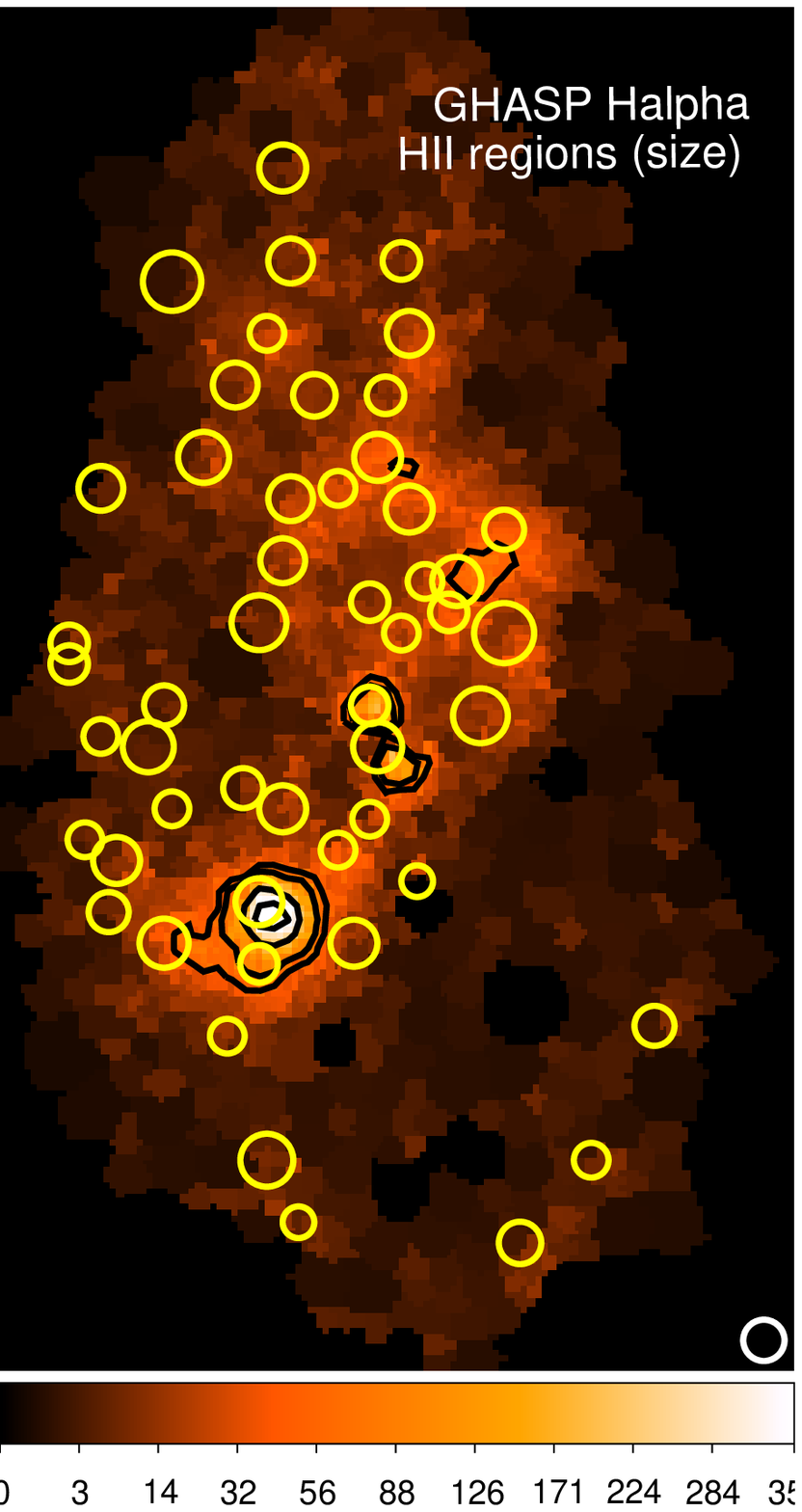} 
\begin{center}
 (c)
\end{center}
    \end{minipage}
        \begin{minipage}{0.2\textwidth}
        \centering
		\includegraphics[width=1\textwidth, angle=0]{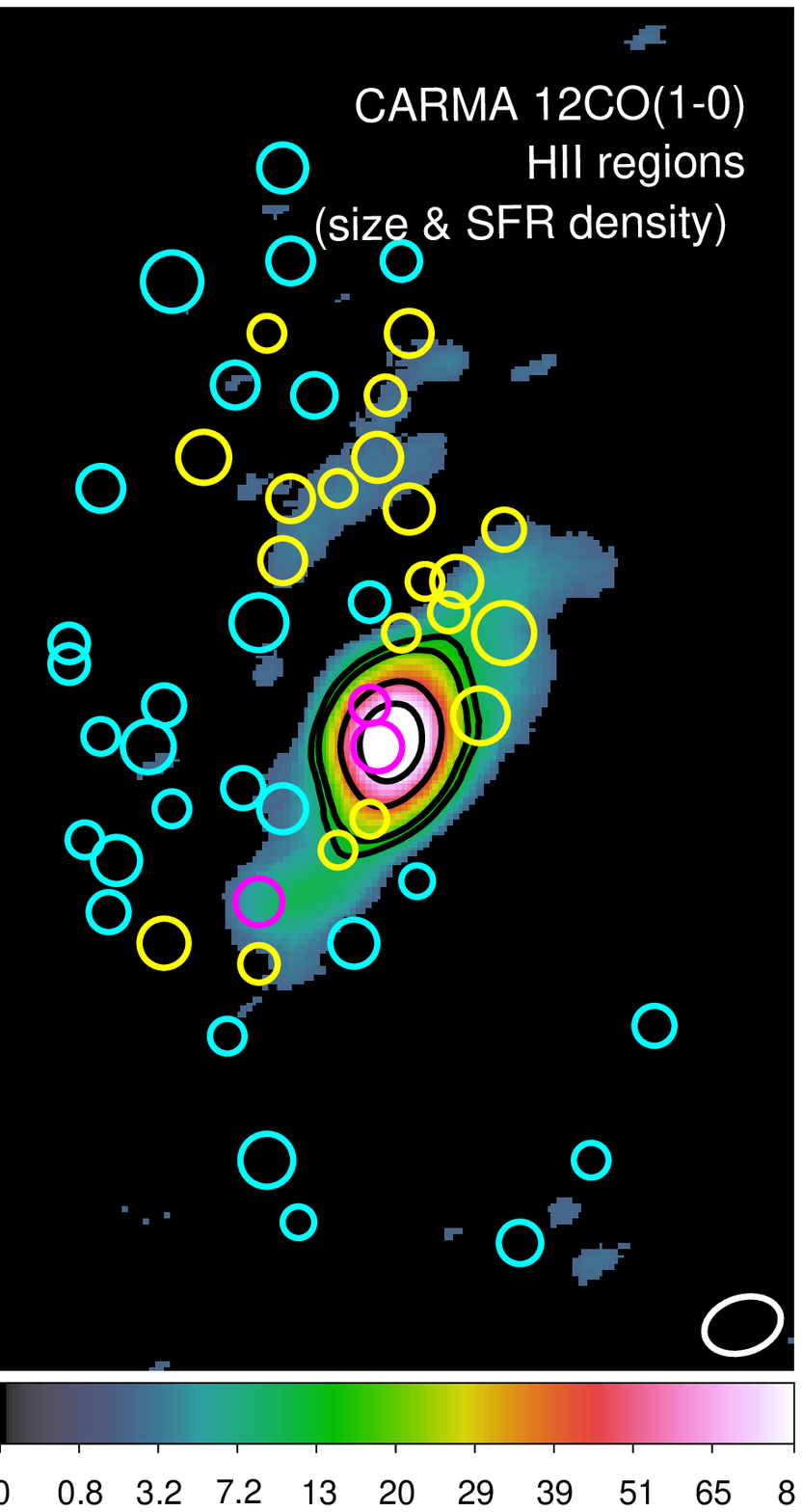} 
\begin{center}
 (d)
\end{center}
    \end{minipage}
    	\end{center}
	\caption{Star formation tracers of NGC 5430. Multi-wavelength  intensity maps are shown in color scale and contours, where contours are plotted with steps of 10, 15, 40, and 70 \% of the maximum intensity.  Resolution of the observations is shown in the lower-right corner of each panel.
(a) Radio continuum in 20 cm (1.4 GHz) observed by VLA \citep{Con98}.
(b) H$\alpha$ image from the GHASP survey \citep{Epi08}.  (c) HII regions  (circles) identified by \citet{Bri12} overlaid on panel (b). Size of circles denote the physical size of HII regions.  (d) HII regions (circles) of  \citet{Bri12} overlaid on $^{12}$CO image. The size and color of circles indicate the physical size and star formation rate density ($\Sigma_{\mathrm{SFR}}$) of HII regions, respectively.
The colors denote, magenta: $\Sigma_{\mathrm{SFR}}$ $>$ 0.05 M$_{\solar}$ yr$^{-1}$ kpc$^{-2}$, yellow:  0.005 $<$ $\Sigma_{\mathrm{SFR}}$ $<$ 0.05 M$_{\solar}$ yr$^{-1}$ kpc$^{-2}$, and cyan: $\Sigma_{\mathrm{SFR}}$ $<$ 0.005 M$_{\solar}$ yr$^{-1}$ kpc$^{-2}$.}
	\label{fig_stellar}
\end{figure*}

History of star formation can be quantified by using star formation rates (SFRs) tracing different timescales. In this work, we used the SFRs estimated from the infrared (SFR(IR)), radio continuum (Figure \ref{fig_stellar}(a), SFR(RC)), and H$\alpha$ (Figure \ref{fig_stellar}(b) and \ref{fig_stellar}(c), SFR(H$\alpha$))  to constrain the history of star formation. The three wavelengths respectively trace the average SFRs during the last $\sim$ 100 Myr, several 10s of Myr, and several Myr, respectively  \citep{Ken98Rev,Mur11,Cal13}. All SFRs in this work were estimated based on the Kroupa initial mass function (IMF). 

Infrared observations suggest that the average SFR during the last $\sim$ 100 Myr is significantly higher than that of the Milky Way. Infrared emission originates from the dust heated by stars younger than  10$^{8}$ yr; therefore, SFR(IR) traces a relatively long history of star formation. The luminosity-to-SFR relation, calibrated by \citet{Cal13}, is: 
\begin{equation}
\frac{\mathrm{SFR}}{[\mathrm{M_{\solar}\: yr^{-1}}]}=2.84\times 10^{-44}\frac{L_{\mathrm{IR}}}{[\mathrm{erg\;s^{-1}}]},
\label{eq_sfr_ir}
\end{equation}
where $L_{\mathrm{IR}}$ is the infrared bolometric luminosity in the 5 to 1000 $\mu$m range. The $L_{\mathrm{IR}}$ of NGC 5430 is  10$^{10.88}$ $L_{\solar}$  as obtained from IRAS, resulting in the SFR(IR) of $\sim$ 8 M$_{\solar}$ yr$^{-1}$, $\sim$ 8 times higher than that of our galaxy. 

The value of the SFR(RC) suggests that an instantaneous starburst had occurred  $\sim$ 10 Myr ago. Radio continuum at 1.4 GHz is mainly the non-thermal synchrotron radiation ($\sim$ 90\%) produced by relativistic electrons ($S$). These electrons are accelerated by the supernovae of massive stars. Therefore, radio continuum traces the star formation history on the timescale corresponding to the lifetime of the lowest massive stars (8 M$_{\solar}$) that could become supernovae, which is  $\sim$ 40 Myr. \citet{Con02} and  \citet{Hes14} suggested a derivation of SFR based on the radio continuum: 
\begin{equation}
\frac{\mathrm{SFR}}{[\mathrm{M_{\solar}\: yr^{-1}}]}=0.75\times 10^{-21}\frac{L_{\mathrm{1.4\, GHz}}}{[\mathrm{W\, Hz^{-1}}]},
\label{eq_sfr_global_rc}
\end{equation}
where $L_{\mathrm{1.4\, GHz}}$ is the luminosity in at 1.4 GHz.
A small fraction of radio continuum  originates from free-free emission ($S_{\mathrm{th}}$).
We use the  following relation:
\begin{equation}
\frac{S}{S_{\mathrm{th}}}=1+10\left ( \frac{\nu }{\mathrm{GHz}} \right )^{0.1-\alpha }
\end{equation}
to subtract the thermal emission from the total flux, where $\alpha$ is the non-thermal spectral index and $\nu$ is the observed frequency.
We adopt a typical value of $\alpha$ $\approx$ 0.8 \citep{Con02}.
The total flux at 1.4 GHz measured by the VLA  was  65.9 mJy  \citep{Con98}.
After subtracting $S_{\mathrm{th}}$ from $S_{\mathrm{1.4GHz}}$, $L_{\mathrm{1.4GHz}}$  becomes 1.3 $\times$ 10$^{22}$ W Hz$^{-1}$.  The corresponding SFR is  $\sim$ 10 M$_{\solar}$ yr$^{-1}$, one order of magnitude higher than that of the Milky Way. Among the global SFR(RC), $\sim$ 1 M$_{\solar}$ yr$^{-1}$  is contributed by the W-R region. The instantaneous starburst may be a result of the satellite collision. Simulations predict that the highest SFR occurs during the final merging phase of two galaxies and during the post-merging phase \citep{Cox08,Lot10}. 

The derived SFR(RC) is  a lower limit owing to the mass loss of massive stars in the W-R region. \citet{Keel82} argued that the mass loss of massive stars in the W-R region is too extreme, so that only $\sim$ 4\% of the massive stars could become supernovae and contribute to the radio continuum. If true, the SFR(RC) is considerably underestimated. If we naively scale the observed SFR(RC) by the fraction of stars that could not become supernovae, the real global SFR during the past $\sim$ 10 Myr will be around 10 (the observed global SFR(RC)) $+$ 1 (the observed SFR(RC) in the W-R region)/0.04 $=$ 35 M$_{\solar}$ yr$^{-1}$, and $\sim$ 25 M$_{\solar}$ yr$^{-1}$ will be contributed by the W-R region.

In contrast to the past, the recent SFR (few Myr) indicated by the total luminosity of the ionized gas is not particularly dramatic, comparable to that of the Milky Way. The H$\alpha$ from the ionized gas traces the most recent star formation owing to the short lifetime of massive stars. Figure \ref{fig_stellar}(b)  shows the H$\alpha$ image obtained from the Observatoire Haute-Provence 1.92-m-diameter telescope as part of the kinematical 3D Gassendi H$\alpha$ survey of spirals survey (GHASP). The H$\alpha$ luminosity ($L_{\mathrm{H\alpha}}$) suggested by the survey data is  2.4 $\times$ 10$^{41}$ erg s$^{-1}$ \citep{Epi08}. SFR(H$\alpha$) is derived from the luminosity-SFR calibration of \citet{Cal13}: 
\begin{equation}
\frac{\mathrm{SFR}}{[\mathrm{M_{\solar}\: yr^{-1}}]}=5.5\times 10^{-42}\frac{L_\mathrm{H\alpha}}{[\mathrm{erg\, s^{-1}}]}. 
\label{eq_sfr_ha}
\end{equation}
The derived global SFR(H$\alpha$) is 1.3 M$_{\solar}$ yr$^{-1}$.
This value suggests that the recent global SFR is  comparable to that of  the Milky Way.

The Galactic level SFR is confirmed by adding up the local extinction-corrected SFR of individual HII regions identified by other independent observations. Figure \ref{fig_stellar}(c)  shows the HII regions identified by  \citet{Bri12} by using the imaging Fourier transform spectrograph SpIOMM on the 1.6-m-diameter Ritchey-Chretien telescopes. The total SFR in the HII regions is 2.4 M$_{\solar}$ yr$^{-1}$, including $\sim$ 0.6 M$_{\solar}$ yr$^{-1}$ from the central region and 1 M$_{\solar}$ yr$^{-1}$  from the W-R region. The values are higher than the SFRs suggested by GHASP observations, probably owing to the uncertainty of the indirect flux calibration in the GHASP survey  (see \cite{Epi08} for the details). In spite of the difference, both global SFR(H$\alpha$) are significantly lower than the global SFR(IR) and SFR(RC). Moreover, the SFR in the W-R region decreases by more than 10 times between the actual SFR(RC) and SFR(H$\alpha$).

NGC 5430 is experiencing the end of the recent W-R starburst. Evolutionary stellar synthesis studies suggest that the age of the W-R region in NGC 5430 is $\geq$ 5.3 Myr \citep{Fer04}. It is at the relatively late stage because W-R starbursts typically fade out within 10 Myr \citep{Sch98}. Our derived SFRs also suggest that the star formation activity had peaked around $\sim$ 35 M$_{\solar}$ yr$^{-1}$, and is rapidly decreasing to the Milky Way value (1 -- 2 M$_{\solar}$ yr$^{-1}$). Moreover, theoretical models suggest that the ionized flux generated by massive stars decreases by 1 -- 2 orders at the end of a W-R starburst  \citep{Sch98}. The difference between the SFRs of actual SFR(RC) and SFR(H$\alpha$) from the W-R region is consistent with this theoretical prediction. 

Finally, if the W-R episode occurred immediately after the satellite collision, the collision time should be between  $\sim$ 5.3 and 10 Myr ago because the W-R period (the time during which the ionized flux decreases to the pre-starburst level or the duration of the appearance of the W-R stars) theoretically does not exceed 10 Myr  \citep{Sch98}.

\subsubsection{Relation Between the Current Star Formation Activity and Molecular Gas}
\label{sec_sfe}

 In this section, we compare  star formation activity  and molecular gas. Since the  H$\alpha$ flux is not  spatially calibrated on  the   image of the GHASP survey (Figure \ref{fig_stellar}(b)) due to the  strategy of the survey, we used the calibrated flux of   HII regions from  \citet{Bri12} and the $^{12}$CO flux of the beam at the same positions for the analyses in this section. Typical size of HII regions is 1 -- 2.5 kpc$^{2}$, comparable to our beam area of $\sim$ 1.5  kpc$^{2}$, but we suggest that the   results still have to be treated with care.

We found that the $^{12}$CO distribution of NGC 5430 is correlated with star formation rate surface density. Figure \ref{fig_stellar}(d) compares the HII regions of  \citet{Bri12} (circles) and our $^{12}$CO image (color scale). Physical areas of HII regions are indicated by the circles sizes. To fairly compare the star formation activity, we normalized the SFRs of  HII regions by the their areas, to obtain the quantity that is known as the star formation rate surface density ($\Sigma_{\mathrm{SFR}}$), measured in the units of  M$_{\solar}$ yr$^{-1}$ kpc$^{-2}$. The $\Sigma_{\mathrm{SFR}}$ of the HII regions is indicated with colored circles. Cyan, yellow and magenta colors denote the $\Sigma_{\mathrm{SFR}}$ below 0.005 M$_{\solar}$ yr$^{-1}$ kpc$^{-2}$, in the 0.005 -- 0.05 M$_{\solar}$ yr$^{-1}$ kpc$^{-2}$ range, and above 0.05 M$_{\solar}$ yr$^{-1}$ kpc$^{-2}$, respectively. The $^{12}$CO emissions are mostly associated with moderate and high $\Sigma_{\mathrm{SFR}}$ (yellow and magenta circles). The missing $^{12}$CO in the low $\Sigma_{\mathrm{SFR}}$ regions can be explained by insufficient sensitivity, by assuming the ratio of $\Sigma_{\mathrm{SFR}}$ to $^{12}$CO flux to be the same in the moderate- and low-$\Sigma_{\mathrm{SFR}}$ HII regions (in other words, the same star formation efficiency). 

Current global star formation efficiency (SFE) of NGC 5430 is consistent with those of other disk galaxies. The current SFE is quantified as SFR(H$\alpha$)/M$_{\mathrm{H_{2}}}$, where M$_{\mathrm{H_{2}}}$ is the molecular gas traced by the $^{12}$CO. The total SFR(H$\alpha$) in the $^{12}$CO region is $\sim$ 2.2 M$_{\solar}$ yr$^{-1}$ obtained by using the measurements of \citet{Bri12}. The total M$_{\mathrm{H_{2}}}$ in our map is $\sim$ 4.7 $\times$ 10$^{9}$ M$_{\solar}$, for the total flux of 5.7 $\times$ 10$^{4}$ Jy beam$^{-1}$ km s$^{-1}$ and the CO-to-H$_{2}$ conversion factor of 2 $\times$ 10$^{20}$ cm$^{-2}$ (K km s$^{-1}$)$^{-1}$. The global SFE of the galaxy is therefore $\sim$ 4.6 $\times$ 10$^{-10}$ yr$^{-1}$. The value is consistent with those for the nearby disk galaxies \citep{Ken98,Big08,Hua15}.

The local SFE exhibits large regional variation. The SFE at the galactic center is $\sim$ 9 $\times$ 10$^{-10}$ yr$^{-1}$. The SFE is calculated by using the total SFR(H$\alpha$) of the central two HII regions from \citet{Bri12} and total $^{12}$CO  flux of the beams associated with them. We found that the SFE is slightly lower compared with other circumunuclear regions, for which the SFEs are typically on the order of 10$^{-9}$ yr$^{-1}$  (e.g., \cite{Ken98}).

The SFE of the W-R region is 10 times higher than the global SFE and the SFE at the galactic center, and is $\sim$ 4.5 $\times$ 10$^{-9}$ yr$^{-1}$, as obtained by using the SFR(H$\alpha$) of the giant HII region at the W-R region and the $^{12}$CO flux of the beam associated with it. The large beam of CO observations may include molecular gas irrelevant to the starburst. Therefore, the derived SFE is a lower limit.

\subsubsection{The Externally Triggered Starburst at the Bar End}

\citet{Wat11} suggested that the enhancement of star formation efficiency at the bar ends is owing to the effective cloud-cloud collisions triggering the star formation. This questions whether HOLM 569B is indispensably required to explain the W-R starburst at the eastern bar end. W-R stars do appear at the bar end of the Milky Way and are more concentrated towards the spiral arms, suggesting a possible promotion of massive star formation via galactic structures  (e.g., \cite{Dav12}). The western bar end is lacking unusual star formation. Because the W-R starburst is transient, it is possible that it had taken place and terminated at the opposite side end. In other words, the two bar sides do not evolve simultaneously \citep{Keel82}. This is certainly a possible alternative. 

Nonetheless, the W-R region exhibits several star formation features that do not favor an internal origin. The surface density of the W-R stars in the W-R region of NGC 5430 is more than 100 times higher than that of the Milky Way \citep{Mas03,Mau11,Dav12,Del13}. Moreover, the HII region associated with the W-R region deviates from the global HII regions in terms of the luminosity function, which is similar to our Milky Way  \citep{Keel82}. Such a distortion of the HII regions luminosity function is not observed in the internally triggered extreme starburst environments such as circumnuclear starbursts (e.g., \cite{Ken89,Fei97,Alo01}). Finally, the star forming region in the W-R region is particularly young ($\sim$ 5 Myr) compared with the rest of the bar area, which is 10 -- 12 Myr old. All these features suggest that the W-R region has been likely formed by the collision with HOLM 569B.

\subsubsection{Dense Gas Content and Future Star Formation Activity}
Dense gas content ($>$ 10$^{4}$ cm$^{-3}$) of molecular gas can be estimated in terms of dense gas fraction.Table \ref{TAB_line_ratios}  lists the intensity of dense gas HCN (1-–0) and the ratio of HCN/$^{12}$CO (1--0) ($R_{\mathrm{dense}}$) in the central and W-R regions. $R_{\mathrm{dense}}$ increases from 0.01 -- 0.06 in non-starburst nuclei to 0.06 -- 0.2 in starburst nuclei \citep{Mat10}. $R_{\mathrm{dense}}$ in the central region of NGC 5430 is within the range of non-starburst nuclei, 0.018. The value for the W-R region remains undetermined, $<$ 0.070. $R_{\mathrm{dense}}$ is often used to infer the fraction of dense gas ($f_{\mathrm{dense}}$) because the critical density of HCN (1-–0) is as high as $10^{4-5}$ cm$^{-3}$. Together with the adopted $X_{\mathrm{CO}}$ and the HCN-to-H$_{2}$ conversion factor ($X_{\mathrm{HCN}}$) of 1.3 $\times$ 10$^{21}$ cm$^{-2}$ (K km s$^{-1}$)$^{-1}$ \citep{Sol92}, $R_{\mathrm{dense}}$ $=$ 0.018 corresponds to $f_{\mathrm{dense}}$ as low as $\sim$ 12 \%. In the W-R region, $f_{\mathrm{dense}}$ is nearly four times higher than that in the central region (46\%), by using the upper limit of 0.070 on $R_{\mathrm{dense}}$. 

The dense gas fraction in the central region of NGC 5430 is lower than that in the circumnuclear regions of other strong-barred galaxies ($\sim$ 50 -- 100\%, e.g., \cite{Koh99,Pan13}), implying that even though the strong bar is efficiently transporting molecular gas to the center, the majority of the gas does not become dense gas. The low density gas fraction explains the above-mentioned low SFE of the central region compared with other circumnuclear regions. According to the relation of HCN luminosity to SFR that was suggested by \citet{Gao04}, the future SFR formed by this dense gas will be $\sim$ 0.4 M$_{\solar}$ yr$^{-1}$, comparable to the current SFR in the central region indicated by H$\alpha$.

The future SFR, arising from the dense gas in the W-R region, will be at most 0.15 M$_{\solar}$ yr$^{-1}$, nearly 10 times lower than the current SFR in the W-R region, which is $\sim$ 1 M$_{\solar}$ yr$^{-1}$. This is consistent with the previous discussion suggesting that the star formation activity in the W-R region is decaying and will continue to decrease in the future.

\section{Initial Mass Ratio}
\label{sec_mass_ratio}
We  roughly estimate the initial mass ratio of NGC 5430 and its companion based on the observed morphology, velocity, and star formation properties.  Major mergers can destroy galaxy disks and produce remnants with properties similar to those of observed giant ellipticals (but some studies have reported the possibility of formation of a disk galaxy in a major merger (e.g., \cite{Spr05}). We rule out this possibility since the candidate of  companion has been suggested to be embedded in NGC 5430 \citep{Keel82}.
Intermediate-mass mergers (4:1 -- 7:1) tend to produce both   ellipticals and hybrid systems, which are disk-like galaxies  supported by velocity dispersion rather than rotation \citep{Bou04,Bou05}.  The value of rotation velocity to velocity dispersion of these hybrid systems ($V_{\mathrm{rot}}/\sigma$) is  small, 0 $<$ $V_{\mathrm{rot}}/\sigma$ $\leq$ 2 \citep{Jog02,Naa03,Bou04,Bou05}.
The maximum H$\alpha$ velocity dispersion of the galaxy is around 70 km s$^{-1}$, and the deprojected H$\alpha$ rotational velocity in the galactic disk is $\sim$ 300 km s$^{-1}$ (corrected with an inclination of 32$^{\circ}$ from \cite{Epi08}), resulting to a $V_{\mathrm{rot}}/\sigma$ of $\sim$ 4. Hence, NGC 5430 is strongly rotationally-supported as a regular disk galaxy.   
Therefore the initial mass ratio between NGC 5430 and its satellite is likely larger than 7:1.
Simulations show that, for a primary galaxy with a stellar mass of NGC 5430 value\footnote{Stellar mass of NGC 5430 is $\sim$ 5.3 $\times$ 10$^{10}$M$_{\solar}$ from the Spitzer S$^{4}$G project \citep{She10}.},  enhanced star formation only occurs in pairs with mass ratio $<$ 20:1  \citep{Cox08,Lot10}.
Therefore the initial mass ratio of NGC 5430 is likely between 7:1 -- 20:1, and probably around 10:1 to generate a star formation timescale (inverse of SFE) of $<$ 2 Gyr at the initial coalescence phase \citep{Cox08}.

\section{Summary}
\label{sec_summary}
In this paper, we constrain the minor-merging and star formation history  of the three-spiral Wolf-Rayet (W-R) galaxy NGC 5430 and its satellite HOLM 569B embedding in the eastern bar of NGC 5430.

Molecular gas tracers $^{12}$CO (1--0) and $^{13}$CO (1--0) were observed by CARMA with a resolution of $\sim$ 1.3 kpc (\S\ref{sec_obs}). The observed area is $\sim$ 16 kpc.
The main results are (\S\ref{sec_results}):
 \begin{itemize}
  \item $^{12}$CO (1--0) is confidently detected in the central region, entire bar, and an asymmetric spiral arm (the third spiral)  of NGC 5430. Solid detections of $^{13}$CO (1--0) show  at the central region and the eastern bar with W-R region. 
  \item The molecular bar shows as offset ridges as seen in typical barred galaxies. The offset ridges of the two sides of the bar are symmetric around the galactic central region, but the eastern bar are stronger in both CO lines.
  \item  NGC 5430 has a strong bar suggested by the high central concentration of molecular gas and the high ellipticity of the bar.  
  \item Velocity field of $^{12}$CO (1--0)  matches the velocity field and the Position-Velocity diagrams of H$\alpha$, which was observed with higher spatial and velocity resolutions.
  \item Rotation curve of H$\alpha$  shows no obvious distortion by galaxies merger. However,  inhomogeneous mass distribution is suggested by a small bump of rotation curve around the W-R region, implying that the merger system has nearly, but not fully, stablized yet. The outer rotation curve are more asymmetric between the two sides of the galaxy due to either the past interaction with HOLM 569B  or/and ongoing interaction with members in the galaxy group. 
  \item Since the minor merger systems kinematicaly stablize after 1 -- 2 Gyr after the first encounter, NGC 5430 likely captures HOLM 569B on this timescale.
\end{itemize}
 
Physical properties of molecular gas at the central region and the W-R region are inferred by the Large Velocity Gradient calculations (LVG) and the  line ratios of $^{12}$CO (1--0)/$^{13}$CO (1-0) ($R_{1-0}$),  $^{12}$CO (2--1)/$^{12}$CO (1--0) ($R_\mathrm{12CO}$), and $^{12}$CO (2--1)/$^{13}$CO (2--1) ($R_{2-1}$) (\S\ref{sec_phy_prop_sf}). The later two ratios are cited from literature. The ratio of $R_{2-1}$ is only available in the central region. 
The main results are:
 \begin{itemize}
  \item The line ratios of the central region are $R_{1-0}$ $\approx$ 10, $R_\mathrm{12CO}$ $\approx$ 0.7, and $R_{2-1}$ $\approx$ 12.6. The ratios are in the range of starburst regions, but close to boundaries between starburst and the relatively quiescent molecular regions (e.g., the Galactic disk molecular gas). The ratios of the W-R region are consistent to the galactic disk molecular gas, with $R_{1-0}$  and $R_\mathrm{12CO}$ and $\sim$ 6 and 0.54, respectively.
    \item Temperature of the central region derived from the LVG calculations is 10 –- 40 K, warmer than the W-R region, which is 10 K.  These results imply that the W-R region is  small and compact, thus the  bulk molecular gas measured by low-resolution observations includes significant amount of  low-temperature molecular gas irrelevant to the W-R starburst. In contrast, in spite of a lower degree of star formation, star formation  is more spatially extended in the central region, leading to a higher derived temperature.
    \item Gas density is 200 -- 1600 cm$^{-3}$ in both regions, consistent to the critical density of CO lines. 
\end{itemize}

Star formation history is constrained using star formation rates (SFRs) that trace different timescales, including infrared (SFR(IR)) for the average of past 100 Myr, radio continuum (SFR(RC)) for the past a few times 10 Myr, and H$\alpha$ (SFR(H$\alpha$)) for the past a few times 1 Myr (\S\ref{sec_phy_prop_sf}).
The main results are:
 \begin{itemize}
   \item The observed global SFR(IR), SFR(RC) and SFR(H$\alpha$) are $\sim$ 8, 10, and 2.4 M$_{\solar}$ yr$^{-1}$, respectively. The observed SFR(RC) might be an lower limit due to the mass loss of massive stars. The \emph{real}  SFR(RC) is probably as high as $\sim$ 35 M$_{\solar}$ yr$^{-1}$. These SFRs suggest that there is an instantaneous starburst around a few times 10 Myr ago. Together with the previous   stellar synthesis study of the W-R region and the theoretical model of W-R starburst, we suggest that the starburst started around 5.3 -- 10 Myr ago, likely triggered by the collision of the satellite.
   \item We found that the current star formation efficiency of the  central region of NGC 5430 is  low compared to other  galaxies. Moreover, the current dense gas fraction of the central region  inferred by the ratio of HCN/$^{12}$CO (1--0) is also low, implying a low SFR in the future as well.
   \item  Current SFE of the W-R region is 10 times higher than the central region. However, SFR of the W-R region is rapidly decreasing by more than 10 times from the collision time to the present time. The future SFR implied by the amout of dense gas suggests that the SFR of the W-R region will continue to decrease.
\end{itemize}

We  discussed the possible initial mass ratio of NGC 5430 and its satellite to be 7:1 -- 20:1 based on the observed morphology, velocity, and star formation properties (\S\ref{sec_mass_ratio}). 

Finally we note that  theoretical simulation is still necessary to constrain the merger history of NGC 5430 system.
Each systems may have  unique satellite orbits that would significantly affect their current morphology and star formation activity.
Models of NGC 5430 and the direct comparison of the multi-wavelength observations will yield important constraints to the this system.
Future high resolution HI observations would be valuable to reveal large scale dynamic structures beyond the molecular and stellar disk and provide a  knowledge of the possible interaction orbits.
Wide-field  CO observations are also indispensable to reveal the gas distribution and gas supply to star formation in detail, as well as the possible influence of galaxies collisions towards the central region, e.g, the asymmetric SFE at the contact point of the circumnuclear ring.

\vspace{15pt}

The authors thank the anonymous referee for the careful reading of the manuscript and  comments. HAP and MU thank the instructors and the local organizing committee of CARMA summer school 2013 for the help of  observations and the  organization of the summer school. Authors thank Jin Koda to provide the code of LVG calculations.  Support for CARMA construction was derived from the states of California, Illinois, and Maryland, the James S. McDonnell Foundation, the Gordon and Betty Moore Foundation, the Kenneth T. and Eileen L. Norris Foundation, the University of Chicago, the Associates of the California Institute of Technology, and the National Science Foundation. Ongoing CARMA development and operations are supported by the National Science Foundation under a cooperative agreement, and by the CARMA partner universities.
This research has made use of the Fabry Perot database, operated at CeSAM/LAM, Marseille, France. This research has made use of the NASA/IPAC Extragalactic Database (NED) which is operated by the Jet Propulsion Laboratory, California Institute of Technology, under contract with the National Aeronautics and Space Administration.


\end{document}